\documentclass{elsarticle}
\usepackage{graphicx}
\usepackage{amsmath}
\usepackage{color}
\usepackage{float}

\begin{document}

\title{Feasible quantum engineering of quantum multiphoton superpositions}

\author{Magdalena Stobi\'nska}
\ead{magdalena.stobinska@gmail.com}

\address{Institute of Theoretical Physics and Astrophysics, University of Gda\'nsk, ul. Wita Stwosza 57, 80-952 Gda\'nsk, Poland}

\address{Institute of Physics, Polish Academy of Sciences, Al.\ Lotnik\'ow 32/46, 02-668 Warsaw, Poland}

\begin{abstract}
We examine an experimental setup implementing a family of quantum non-Gaussian filters. The filters can be applied to an arbitrary two-mode input state. We assume realistic photodetection in the filtering process and explore two different models of inefficient detection: a beam splitter of a small reflectivity located in front of a perfect detector and a Weierstrass transform applied to the unperturbed measurement outcomes. We explicitly give an operator which describes the {\it coherent} action of the filters in the realistic experimental conditions. The filtered states may find applications in quantum metrology, quantum communication and other quantum tasks.
\end{abstract}

\maketitle

\section{Introduction}

Recent technological advances in the field of quantum optics, such as integrated optics schemes, allow unprecedented control of various degrees of freedom of  optical quantum systems. Nevertheless, generation of quantum states of light beyond the set of squeezed vacuum states (deterministic) and, to some approximation, pairs of entangled photons (probabilistic and in a postselective way), still remains challenging. Most of the protocols implementing quantum technologies require however the use of more complex states. They often belong to the class of non-Gaussian quantum states (states with non-Gaussian quasi-probability distribution~\cite{Scully}). Their generation seems possible exploiting the efficient source of quantum light based on parametric down conversion (PDC) and quantum engineering.

Quantum engineering implements general quantum operations, often described by the positive operator-valued measures (POVMs). Since Gaussian quantum superpositions of light are produced directly by the PDC source~\cite{ZUK}, it is interesting to implement non-Gaussian operations.  They will turn the Gaussian states into the non-Gaussian  multiphoton quantum superpositions of certain properties, required for realization of concrete quantum tasks. Such states are necessary, for example, for obtaining a quantum speed-up in computation with quantum algorithms~\cite{Gottesman} and quantum super-resolution in quantum phase estimation using the N00N and NmmN states~\cite{QMetrology}. They may find also application in Bell inequality tests performed with homodyne detection, the most easy accessible, fast and efficient photodetector at present. These tests can be used to certify quantum devices~\cite{DIBV}.

Up to date, the most often experimentally realized non-Gaussian operations comprise probabilistic single photon addition~\cite{Bellini} or subtraction~\cite{Grangier}. They closely approximate action of the creation and annihilation operator, respectively. An alternative method of implementing the creation operator, based on repeated spontaneous parametric down-conversion, was theoretically investigated in~\cite{Kwiat}. These operations can alternate, add or cancel certain components of the initial engineered superposition. Since the probability of the success is very small, they cannot be applied iteratively.   Quantum engineering in the form of a quantum filter can only cancel certain components of the initial superposition. Quantum filtering was demonstrated for one- and two-photon Fock states~\cite{Sanaka,Resch}. The filters were based on the Hong-Ou-Mandel interference~\cite{HOM,Cosme,Walborn} and were capable of blocking single photons over photon pairs. A quantum device capable of filtering out two-mode states of light with mode populations differing by more than a certain threshold, was proposed in~\cite{MDF}. It is called the modulus of intensity difference filter (MDF) and is based on the multiphoton Hong-Ou-Mandel interference performed in a feed-forward loop. It allows engineering of the multiphoton quantum superpositions in a way which is preserving specific superpositions. This may turn them useful for Bell test and quantum metrology ~\cite{Bell1,Bell2}. Some of the features of this filter has already been experimentally demonstrated in~\cite{Masha}. 

In this paper we examine the experimental scheme from~\cite{MDF} and show that in fact it implements a whole family of quantum non-Gaussian filters (the MDF is just one of the possibilities). The filters can be applied to an arbitrary input state. We assume realistic photodetection in the filtering process. This is an important step in the analysis of quantum filtering, since lossy detection is detrimental for the possibility of observation of quantum effects. We model the inefficient detection in two different ways: with a beam splitter of a small reflectivity located in front of a perfect detector (the usual way) and a Weierstrass transform, which implements a Gaussian blur on the unaffected measurement outcome distribution.  We show how the filter acts on an input quantum superposition by computing the photon number distributions and purity of the filtered states. We also construct the Kraus operator for the filters which reveals their {\it coherent} action on an arbitrary input. The filtered states may find applications in quantum metrology, quantum communication and other quantum tasks. 

This paper is organized as follows. Section \ref{sec:efficient_filtering} presents the theoretical description of the experimental setup implementing a family of quantum non-Gaussian filters. In subsection \ref{ssec:mHOM} we recall the Hong-Ou-Mandel interference and generalize it to the multiphoton case. In subsection \ref{ssec:filter} we show state evolution within the feed forward loop. In subsection \ref{ssec:examples} we analyze the family of the output states which can result from the filter and we construct its Kraus operator describing the action on an arbitrary two-mode input state. In Section \ref{photodetection} we discuss two models of realistic photodetection. We also give the Kraus operator describing the filters in presence of inefficient detection.  In section \ref{sec:results} we present numerical computations demonstrating the action of the filter with realistic photodetection for two important examples of initial quantum superpositions. The results are given for two population regimes: the few photon and the mesoscopic population of photons in the initial superpositions. The paper is summarized in conclusions.

\section{Experimental setup implementing a family of non-Gaussian filters}
\label{sec:efficient_filtering}

\begin{figure}
  \begin{center}
   \includegraphics[height=6cm]{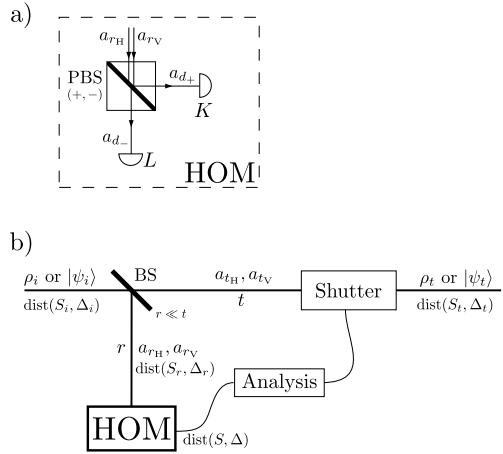}
 \end{center}
  \caption{Experimental setup implementing a family of non-Gaussian quantum filters. Description of the setup is given in the main text.}           
 \label{filter}
\end{figure}

In this section we introduce an experimental setup (Fig.~\ref{filter}b) which implements a family of non-Gaussian quantum filters. These filters preserve the symmetry present in photon number distribution of an input two-mode quantum state. They implement a non-Gaussian operation and prepare the input beam for further quantum tasks. The principle of work of the filters is based on the multiphoton Hong-Ou-Mandel (HOM) interference (Fig.~\ref{filter}a) observed in a feed forward loop. The same setup was used before to implement the modulus of intensity filter discussed in~\cite{MDF}.

\subsection{Multiphoton Hong-Ou-Mandel interference}
\label{ssec:mHOM}

\begin{figure}
  \begin{center}
    \includegraphics[height=3cm]{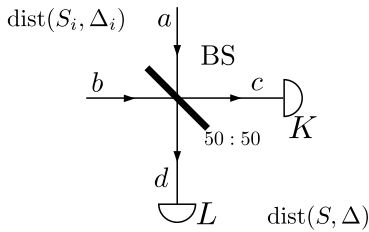}
   \end{center}
  \caption{Experimental setup for observation of the Hong-Ou-Mandel interference. Behind the beam splitter are located photon counting detectors. $\text{dist}(S_{(i)},\Delta_{(i)})$ denotes the probability distribution of the total photon number $S_{(i)}$ and occupation difference $\Delta_{(i)}$ at the output (input) ports of the beam splitter.}           
 \label{HOM}
\end{figure}

\begin{figure}
  \begin{center}
    \includegraphics[height=3.2cm]{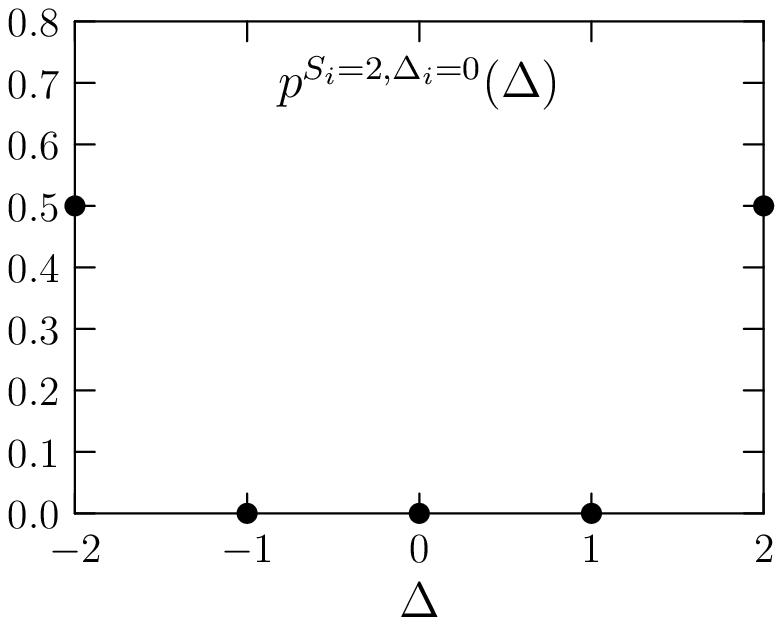}
    \includegraphics[height=3.2cm]{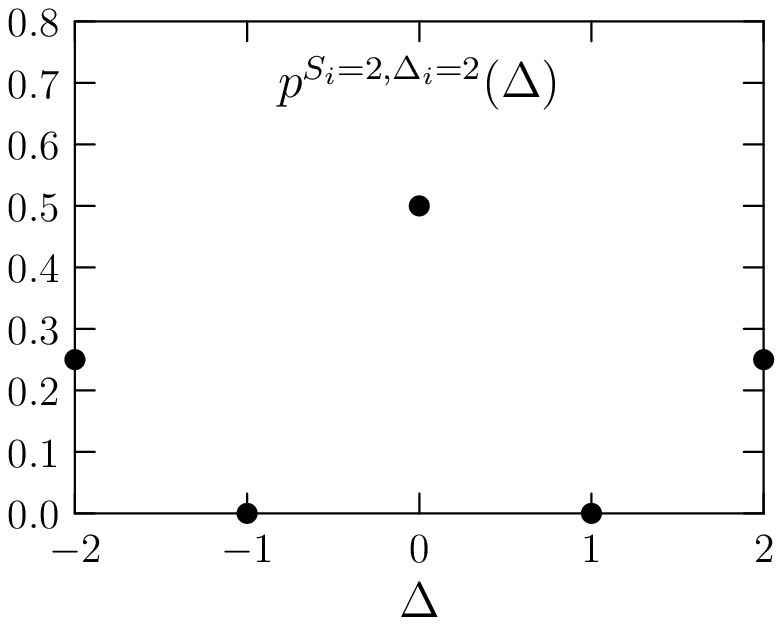}
   \end{center}
  \caption{Distribution of the output population difference at the exit ports of a 50:50 beam splitter for the interference of 2 photons which enter the beam splitter separately -- $p^{S_i=2,\Delta_i=0}(\Delta)$ or via the same input port -- $p^{S_i=2,\Delta_i=2}(\Delta)$. HOM interference manifests itself in the double-peaked shape of $p^{S_i=2,\Delta_i=0}(\Delta)$.}           
 \label{2phHOM}
\end{figure}

Let us recall the two-photon Hong-Ou-Mandel interference. We analyze the experimental setup shown in Fig.~\ref{HOM}.
Two identical photons (one in mode $a$ and the other one in $b$) interfering at a balanced (50:50) beam splitter (BS) always exit together. Behind the beam splitter they are registered by the photon counting detectors. The only possible measurement outcomes are either $K=0$ and $L=2$ or $K=2$ and $L=0$. Thus, the probability distribution of the output population difference between the output ports of the beam splitter denoted by $\Delta=L-K$ is $p(\Delta = \pm 2) = 1/2$ and $p(\Delta = 0) = 0$. In this case the events of ``large'' (equal to the total photon number) output difference are more likely than the events of ``small'' (zero) output difference. If the two photons enter the beam splitter through the same input port (e.g. $a$) and the other port ($b$) is empty, the most likely is that the photons exit separately, i.e. $K=L=1$ and $p(\Delta = 0) = 1/2$. We denote the initial total photon number by $S_i=2$ and the initial population difference by $\Delta_i$. The probability distributions of the output population difference $p^{S_i=2,\Delta_i}(\Delta)$ are displayed in Fig.~\ref{2phHOM} for $\Delta_i=0,2$.

\begin{figure}
  \begin{center}
    \raisebox{4cm}{a)}
    \includegraphics[height=4cm]{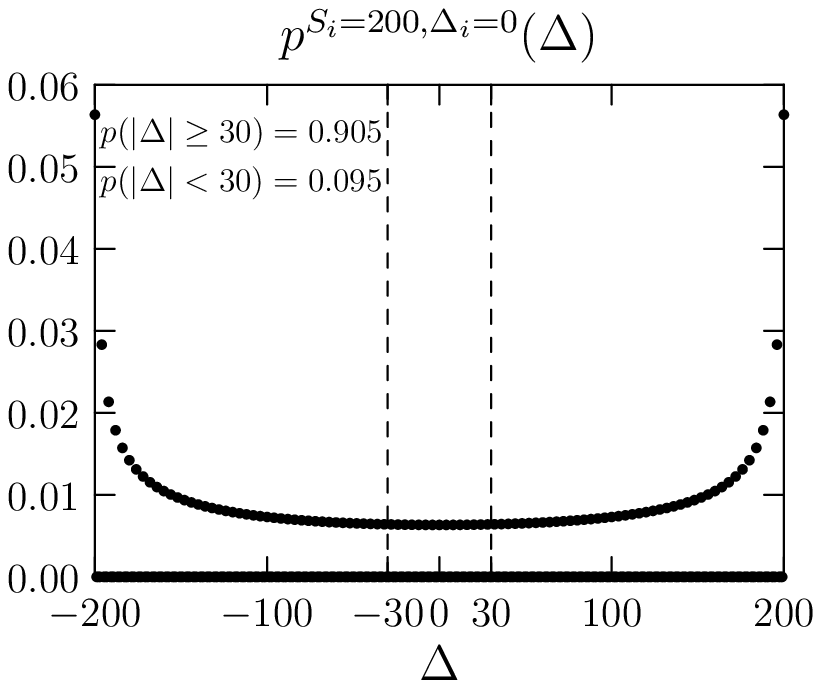}\\
    \raisebox{4cm}{b)}
    \includegraphics[height=4cm]{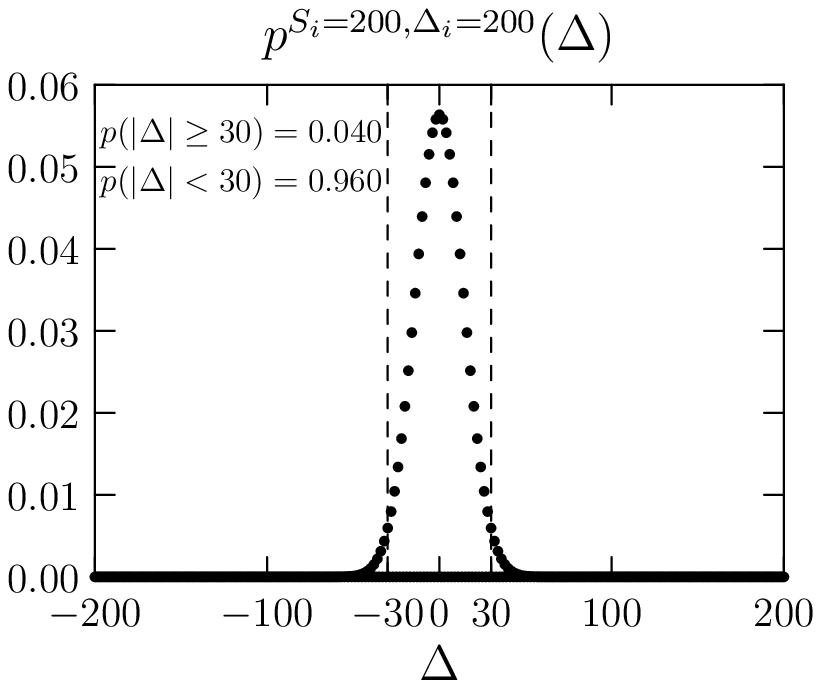}
  \end{center}
\caption{Distributions of the population difference in the output ports of a 50:50 beam splitter $p^{S_i,\Delta_i}(\Delta)$ after interference of a Fock state $\lvert n,m\rangle$, with $n+m=S_i=200$ and with the initial population difference $n-m=\Delta_i = 0$ (a), $\Delta_i = 200$ (b). The distributions are symmetric: $p^{S_i,\Delta_i}(\Delta)=p^{S_i,\Delta_i}(-\Delta)$. The vertical dashed lines show the threshold $\delta_{th}=30$. The probability that $|\Delta| \ge 30$ is given by $p(|\Delta| \ge 30)$.}
  \label{fig:single_mzi_s200}
\end{figure}

Similar effect to the one shown in Fig.~\ref{2phHOM} takes place if higher photon number (Fock) states interfere at the 50:50 beam splitter. If two equal Fock states $\lvert n\rangle$ enter in mode $a$ and $b$, the most likely event is that the output population difference is large ($\Delta = L-K\approx S_i = 2n$)~\cite{Campos}. If a Fock state $\lvert 2n\rangle$ interferes with the vacuum state, most likely the photons will split equally between the output ports ($\Delta \approx 0$). In order to examine this effect in detail, we explicitly derive the form of the output state after interference of two Fock states $|n\rangle$ and $|m\rangle$ on the beam splitter
\begin{align}\label{UBS}
\mathcal{U}_{\text{BS}} \lvert n, m\rangle_{a,b}
&= \tfrac{1}{\sqrt{n!\, m!}} \tfrac{1}{\sqrt{2^{n+m}}}
\sum_{p=0}^n \sum_{q=0}^m \binom{n}{p} \binom{m}{q} \\&\quad\quad\quad
(-1)^{v-p}\sqrt{(p+q)!(n+m-p-q)!}\nonumber\\&\quad\quad\quad
|p+q, v+w-p-q\rangle_{c,d}.\nonumber
\end{align}
The operator $\mathcal{U}_{\text{BS}}$ describes the action of the 50:50 beam splitter on two input modes $a$ and $b$ ($|n\rangle = \tfrac{{a^{\dagger}}^n}{\sqrt{n!}}|0\rangle$, $|m\rangle = \tfrac{{b^{\dagger}}^m}{\sqrt{m!}}|0\rangle$). In the Heisenberg picture it transforms the creation operators in the following way $\mathcal{U}^{\dagger}_{\text{BS}}a^{\dagger} \mathcal{U}_{\text{BS}}=(c^{\dagger} + d^{\dagger})/\sqrt{2}$, $\mathcal{U}^{\dagger}_{\text{BS}}b^{\dagger} \mathcal{U}_{\text{BS}}=(c^{\dagger} - d^{\dagger})/\sqrt{2}$, where $c$ and $d$ denote the modes exiting BS. Next, the output state (\ref{UBS}) is measured by the perfect photon counting detectors located behind the beam splitter. This corresponds to a projection of the state (\ref{UBS}) on some Fock states $|K\rangle_c$, $|L\rangle_d$. The total photon number is conserved by the beam splitter and equals $S_i=n+m=K+L=S$.  The probability distribution of the output population difference $\Delta$ conditioned on the initial population difference $\Delta_i=n-m$ and sum $S_i$ reads
\begin{align}
\label{eq:pKL}
p^{S_i,\Delta_i}(\Delta) &= |\langle K, L| \mathcal{U}_{\text{BS}} \lvert n, m\rangle|^2 \\
&= |\langle \tfrac{S-\Delta}{2}, \tfrac{S+\Delta}{2}| \mathcal{U}_{\text{BS}} \lvert \tfrac{S_i+\Delta_i}{2}, \tfrac{S_i-\Delta_i}{2}\rangle|^2,\nonumber
\end{align}
\begin{align}
p^{S_i,\Delta_i}(\Delta) &=  \frac{(\tfrac{S_i-\Delta}{2})! (\tfrac{S_i+\Delta}{2})!}{ 2^{S_i} (\tfrac{S_i-\Delta_i}{2})! (\tfrac{S_i+\Delta_i}{2})!}
\label{distribution}\\&\quad
\Bigg|\sum_{p=0}^{\tfrac{S_i+\Delta_i}{2}} \sum_{q=0}^{\tfrac{S_i-\Delta_i}{2}} \delta_{p+q,\tfrac{S-\Delta}{2}}
\binom{\tfrac{S_i+\Delta_i}{2}}{p} \binom{\tfrac{S_i-\Delta_i}{2}}{q}
(-1)^{q}\Bigg|^2.
\nonumber
\end{align}
Fig.~\ref{fig:single_mzi_s200} shows the probability distribution of the output population difference (\ref{distribution}) computed for $S_i=200$ and two extreme cases of $\Delta_i = 0$ and $\Delta_i = S_i = 200$.  These figures reveal the essence of the multiphoton HOM interference: for two equal Fock states interfering on a 50:50 beam splitter the most likely event is that all photons will exit together (the probability distribution $p^{S_i,0}(\Delta)$ is double-peaked); for a nonzero Fock state interfering with the vacuum the most likely event is that the photons will split equally between the output ports (the probability $p^{S_i,S_i}(\Delta)$ is given by the single-peaked binomial distribution). Please note, that for any $S_i, \Delta_i$ the distribution $p^{S_i,\Delta_i}(\Delta)$ is symmetric: $p^{S_i,\Delta_i}(\Delta)=p^{S_i,\Delta_i}(-\Delta)$.

Moreover, the probability distribution $p^{S_i,\Delta_i}(\Delta)$ allows to determine the probability that the modulus of the output population difference is greater or smaller than a certain threshold.  In Fig.~\ref{fig:single_mzi_s200} we took the threshold $\delta_{th}=30$. The probability that $|\Delta|\geq 30$ equals $0.905$ for $\Delta_i=0$ and $0.04$ for $\Delta_i=200$.

Please note that $p^{S,\Delta_i}(\Delta)=p^{S,\Delta}(\Delta_i)$ due to the bi-stochastic nature of these quantum probabilities~\cite{Peres}. This means that the analysis of the measurement outcomes of the detectors located behind the BS ($S,\Delta$) allows to forecast the distribution of the initial population difference ($\Delta_i$) in the input Fock states. Therefore, the plot of $p^{S,\Delta}(\Delta_i)$ is identical to the plot of $p^{S,\Delta_i}(\Delta)$ in Fig.~\ref{fig:single_mzi_s200}. This is one of the two key effects exploited by the setup in Fig.~\ref{filter} implementing a family of non-Gaussian filters.

\subsection{Principle of work of a quantum filter in Fig.~\ref{filter}b}
\label{ssec:filter}

Let us now describe the action of the setup in Fig.~\ref{filter}b. We assume that the input state (either mixed $\rho_i$ or pure $\lvert\psi_i\rangle$) is a two-mode quantum state. For concreteness, we assume the modes to be linear polarizations $H$ -- horizontal, $V$ -- vertical.  The input state impinges on a biased beam splitter  with small reflectivity (e.g. 10:90). The reflected beam $r$ is sent to a polarizing beam splitter (PBS), oriented such that it selects the unbiased polarization modes ($a_{d_{+}}, a_{d_{-}}$) with respect to the incoming linear polarizations ($a_{r_H}, a_{r_V}$). In this case, the action of the polarizing beam splitter $\mathcal{U}_{PBS}$ is the same as $\mathcal{U}_{BS}$ ($a_{d_+}=(a_{r_V}+a_{r_H})/\sqrt{2}$, $a_{d_-}=(a_{r_V}-a_{r_H})/\sqrt{2}$). Thus, the experimental situation in Fig.~\ref{filter}a is identical to the one in Fig.~\ref{HOM}. Now, the annihilation operators $a_{r_H}$, $a_{r_V}$, $a_{d_+}$, $a_{d_-}$  play the role of $a$, $b$, $c$, $d$, respectively. The measurement outcomes of the detectors located behind the PBS ($S, \Delta$) reveal the photon number reflected by the 10:90 beam splitter $S_r=S$ and allow to forecast the population difference $\Delta_r$ before PBS ($S_r$ and $\Delta_r$ play the role of $S_i$ and $\Delta_i$ in the discussion from Subsection~\ref{ssec:mHOM}). The only difference is that now the incoming state impinging on the PBS is not a single Fock state but a superposition of those. We will show below that nevertheless the reasoning from Subsection~\ref{ssec:mHOM} still applies. The outcomes $S$ and $\Delta$ parametrize the family of non-Gaussian filters in Fig.~\ref{filter}b because they allow to choose the shape of the probability distribution of $\Delta_r$. Since the reflected $r$ and transmitted $t$ beams in the feed-forward loop are correlated, $S$ and $\Delta$ also allow to estimate the distributions for the total photon number and population difference between the polarization modes in the incoming ($S_i, \Delta_i$) and transmitted ($S_t, \Delta_t$) beams, which are symmetric like those in Fig.~\ref{fig:single_mzi_s200}. This is the second important effect exploited by the filter. It is especially pronounced for larger photon numbers. For example, in case of the 10:90 beam splitter we most often obtain $\Delta_r\approx 0.1\Delta_i$, $\Delta_t\approx 0.9\Delta_i$. Knowing the distributions of $S_t$ and $\Delta_t$, the analysis box checks the probability that they fulfill certain desired condition $\text{C}(S_t,|\Delta_t|)$ (e.g. $|\Delta_t| > S_t / 2$). If the probability is high enough, the box opens the shutter and passes the transmitted beam for further processing.  If the probability is too small, the shutter remains closed and the transmitted beam is blocked.

For the completeness of our discussion, we will now show that the HOM interference effect displayed in Fig.~\ref{fig:single_mzi_s200} is also observed for an input state being a superposition of Fock states. We summarize a detailed computation presented in Appendix B of \cite{MDF}. Let us assume the input state entering setup in Fig.~\ref{filter}b to be $\lvert\psi_i\rangle =\sum_{n,m}\xi_{n m}\lvert n,m\rangle$. After passing the BS and PBS, the state equals
\begin{align}
& \lvert\psi_{dt}\rangle
= \sum_{n,m}\xi_{nm} \sum_{v=0}^n \sum_{w=0}^m
\dfrac{c_v^{(n)}\, c_w^{(m)}}{\sqrt{v!\, w!}}
\tfrac{1}{\sqrt{2^{v+w}}}
\nonumber\\&\quad
\sum_{p=0}^v \sum_{q=0}^w \binom{v}{p} \binom{w}{q} (-1)^{v-p}
\sqrt{(p+q)!\,(v+w-p-q)!}
\nonumber\\&\quad \lvert p+q,v+w-p-q\rangle_d
 \lvert n-v,m-w\rangle_t,
\end{align}
where $c_k^{(n)}=\sqrt{\binom{n}{k}\,r^k\,t^{n-k}}$, $r$ is the reflectivity of the tapping beam splitter and $t=1-r$. The perfect detectors behind the PBS detect two Fock states $|K,L\rangle_d$ and project the state $\lvert\psi_{dt}\rangle$ to 
\begin{align}
\label{bzyk}
\lvert\psi_{t} \rangle &= {}_d\langle K,L\vert\psi_{dt}\rangle,\\
&= \sqrt{\tfrac{K!L!}{2^{K+L}}}
\sum_{n,m}\tilde{\xi}_{nm} \sum_{v=0}^n \sum_{w=0}^m 
\dfrac{c_v^{(n)}\, c_w^{(m)}}{\sqrt{v!\,w!}}\,\delta_{K+L,v+w}
\nonumber\\&\quad
\left[\sum_{p=0}^v \sum_{q=0}^w \binom{v}{p} \binom{w}{q} 
(-1)^{v-p} 
\,\delta_{K,p+q}\right] \lvert n-v,m-w\rangle_t.\nonumber
\label{wz6}
\end{align}
We note that $v+w=S_r=S$ and $v-w=\Delta_r$ whereas $n-v+m-w = S_t$ and $n-v-(m-w)=\Delta_t$. The coefficients $\tilde{\xi}_{nm}$ are renormalized to ensure normalization of $\lvert\psi_{t} \rangle$. We compute the conditional photon number distribution for the transmitted beam
$p^{K,L}(k,l) = |\langle k,l\lvert\psi_t \rangle |^2$. Here, $k$ denotes the photon number in polarization $H$ and $l$ in $V$. After changing the variables $L$ and $K$ so that they  corresponded to the quantities useful for the filtering we obtain $p^{S,\Delta}(S_t,\Delta_t)$ with $S_t = k+l$, $\Delta_t = k-l$
\begin{align}
&p^{S,\Delta}(S_t,\Delta_t) =
\tfrac{1}{2^S}\left(\tfrac{S+\Delta}{2}\right)!\,\left(\tfrac{S-\Delta}{2}\right)! 
\nonumber\\& \quad
\Big(\sum_{n,m} \tilde{\xi}_{nm} \sum_{v=0}^n \sum_{w=0}^m 
\dfrac{c_v^{(n)}\, c_w^{(m)}}{\sqrt{v!\,w!}}
\nonumber\\& \quad
\delta_{S,v+w}\, \delta_{\tfrac{S_t+\Delta_t}{2},n-v}\,\delta_{\tfrac{S_t-\Delta_t}{2},m-w} 
\nonumber\\& \quad
\sum_{p=0}^v \sum_{q=0}^w \binom{v}{p} \binom{w}{q} (-1)^{v-p}
\,\delta_{\tfrac{S-\Delta}{2},p+q}\Big)^2.
\end{align}
Plots of the above probability  distribution computed for a superposition of Fock states and presented in Appendix B of \cite{MDF}, are similar to the distributions in Fig.~\ref{fig:single_mzi_s200} computed for a single Fock state.

\subsection{Examples of quantum non-Gaussian filters}
\label{ssec:examples}

The way the quantum filter in Fig.~\ref{filter}b alters the incoming two-mode quantum state depends on the information the incoming state carries about the distributions of the total photon number and of population difference between its polarization modes ($\text{dist}(S_i,\Delta_i)$). 

We will consider few generic examples illustrating action of the setup in Fig.~\ref{filter}b. Our main tool will be the photon number distributions $p^{S,\Delta}(k,l)$ depicted for the transmitted beam, before the shutter, conditioned on the measurement of $S$ photons on the reflected beam and population difference $\Delta$ on the detectors behind the polarizing beam splitter. We emphasis that these plots will only visualize qualitatively the action of the setup and will show the photon number distributions \emph{before} the selection performed by the analysis box and the shutter according to a certain condition $\text{C}(S_t,|\Delta_t|)$.

\begin{figure}
\begin{center}
  \raisebox{3.6cm}{a)}\includegraphics[width=3.5cm]{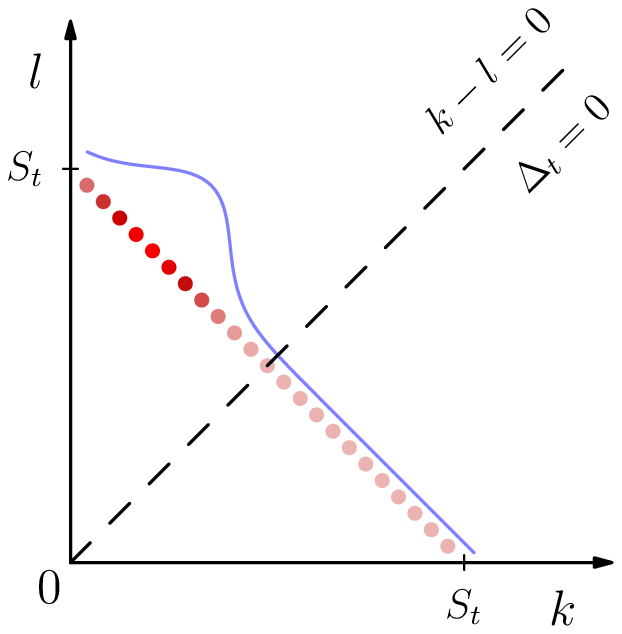}\quad
  \raisebox{3.6cm}{b)}\includegraphics[width=3.5cm]{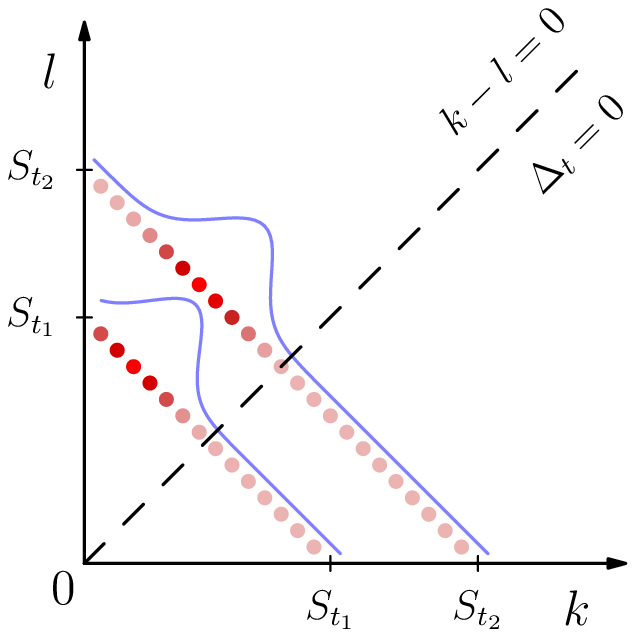}\quad
  \raisebox{3.6cm}{c)}\includegraphics[width=3.5cm]{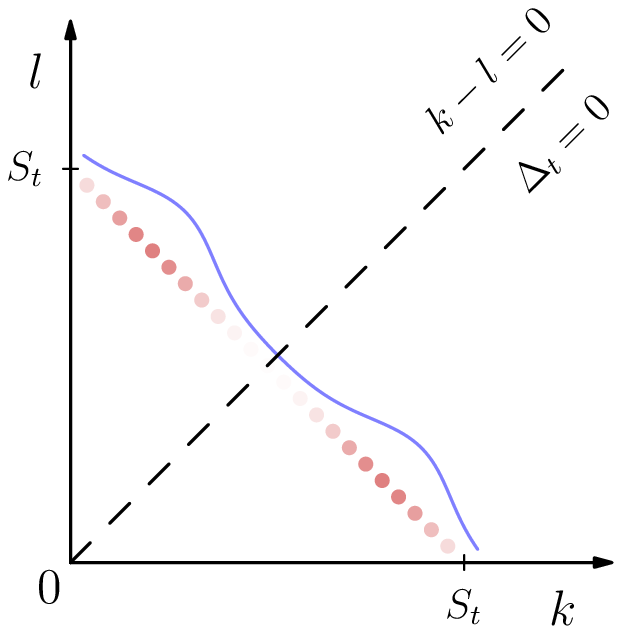}\quad
  \raisebox{3.6cm}{d)}\includegraphics[width=3.5cm]{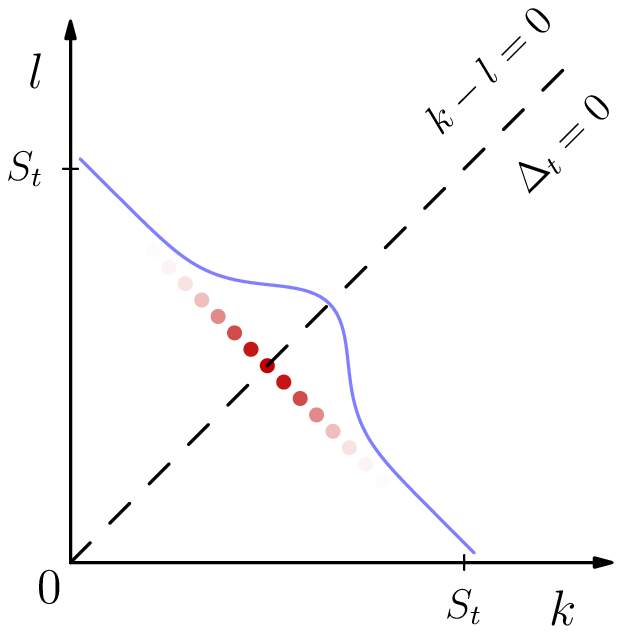}
\end{center}
\caption{Visualization of the photon number distribution for the transmitted beam \emph{before} the shutter in Fig.~\ref{filter}b for different inputs. a) -- a Fock state. In this case $S_t=S_i-S$ is fixed, where $S$ denotes the photon number registered by the detectors on the reflected beam. The filter projects the input state on the line coherently. b) -- a mixture of two Fock states with different photon numbers $S_{i_1}$ and $S_{i_2}$. Here, $S_{t_{1 (2)}}=S_{i_{1 (2)}}-S_{1 (2)}$. The projections on two distinct lines are incoherent with respect to each other. c) \& d) -- superposition of Fock states of equal total photon number and uniform distribution of initial population difference, $S_t=S_i-S$. The projection is also coherent.}
\label{line}
\end{figure}

We start with examples of states where the total photon number is known and well-defined. At first we assume the input state to be a Fock state
\begin{equation}
|\psi_i\rangle = |S_i-N, N\rangle
\end{equation}
with $S_i-N\leq N$ ($\Delta_i\leq 0$).  The measured photon number in the reflected beam gives us precise information of the photon number in the transmitted beam $S_t=S_i-S$. Thus, the only points in the photon number space ($k,l$) of the transmitted beam, for which $p^{S,\Delta}(k,l)$ may be nonzero are those on the line of constant photon number $S_t=k+l$, see Fig.~\ref{line}a.  The distribution is unsymmetric with respect to the line $\Delta_t=0$ (see the blue curve), independently of the values of the measured $S$ and $\Delta$. It indicates that most likely $\Delta_t\leq 0$, revealing the asymmetry in the photon number distribution of the input. From Eq.~(\ref{bzyk}) we notice that $\lvert\psi_t\rangle$ ($\tilde{\xi}_{nm} = \delta_{n,S_i-N}\,\delta_{m,N}$) is a superposition state of components $\lvert N-v,M-w\rangle$. Each component has a fixed photon number $S_t=S_i-S$ and population $\Delta_t=\Delta_i-\Delta_r$ thus, $S_t$ and $\Delta_t$ characterize them completely. In this new notation, the state $\lvert\psi_t\rangle$ equals 
\begin{align}
\lvert\psi_t\rangle ={}& \lvert S_i-S\rangle \sum_{\Delta_r=\Delta_r^{min}}^{\Delta_r^{max}} f^{S,\Delta}_{S_i,\Delta_i}(\Delta_r) \lvert \Delta_i-\Delta_r\rangle,
\label{FockFiltered}
\end{align}
\begin{align}
f^{S,\Delta}_{S_i,\Delta_i}(\Delta_r) &{}= \sqrt{\tfrac{1}{2^S}\tfrac{\left(\tfrac{S+\Delta}{2}\right)!\,\left(\tfrac{S-\Delta}{2}\right)!}{\left(\tfrac{S+\Delta_r}{2}\right)!\,\left(\tfrac{S-\Delta_r}{2}\right)!}}\\
&c_{\tfrac{S+\Delta_r}{2}}^{\left(\tfrac{S_i+\Delta_i}{2}\right)}\, c_{\tfrac{S-\Delta_r}{2}}^{\left(\tfrac{S_i+\Delta_i}{2}\right)}
A^{S}_{\tfrac{S-\Delta}{2}}(\Delta_r),
\end{align}
where $\Delta_r^{min} = \text{min}\{-S,S-S_i+\Delta_i\}$, $\Delta_r^{max} = \text{max}\{S,S_i-S+\Delta_i\}$, $A^{S}_{\tfrac{S-\Delta}{2}}(\Delta_r)$ is given by the square brackets in Eq.~(\ref{bzyk}). The filter projects the Fock input state \emph{coherently} on a line $S_t = k+l$ with $\Delta_t = k-l \in [\Delta_i-\Delta_r^{min}, \Delta_i-\Delta_r^{max}]$. The analysis box and the shutter will further select some components from (\ref{FockFiltered}) according to a condition $\text{C}(S_t,|\Delta_t|)$.  We conclude that the quantum filter performs the following operation on the Fock input states $\lvert \tfrac{S_i + \Delta_i}{2}, \tfrac{S_i - \Delta_i}{2} \rangle \equiv \lvert S_i, \Delta_i\rangle$
\begin{align}
\mathcal{P}^{S,\Delta}_{\text{C}}[S_i,\Delta_i]&= \lvert S_i-S\rangle \langle S_i\rvert \label{P_Fock}\\
&\otimes \left( \sum_{\substack{\Delta_r=\Delta_r^{min}\\\text{C}(|\Delta_r|)}}^{\Delta_r^{max}} f^{S,\Delta}_{S_i,\Delta_i}(\Delta_r) 
\lvert \Delta_i-\Delta_r\rangle \right)\langle \Delta_i\rvert.\nonumber
\end{align} 
If $|\Delta_r^{min}|= |\Delta_r^{max}|$, the function $f^{S,\Delta}_{S_i,\Delta_i}(\Delta_r)$ is symmetric with respect to the line $\Delta_t=\Delta_i$. Note that $f^{S,\Delta}_{S_i,\Delta_i}(\Delta_r)>0$, $\sum_{\Delta_r} f^{S,\Delta}_{S_i,\Delta_i}(\Delta_r) = 1$. The plot of an exemplary $f^{S,\Delta}_{S_i,\Delta_i}(\Delta_r=\Delta_i-\Delta_t)$ is depicted along the line in Figs.~\ref{line}a  -- the blue curve. 

As the next example we consider a mixture of two Fock states
\begin{align}
\rho_i&= q\,|S_{i_1}-N,N\rangle\langle S_{i_1}-N,N|\\
&+ (1-q)\,|S_{i_2}-M,M\rangle\langle S_{i_2}-M,M|. \nonumber
\end{align}
The filter will act independently on each mixture term. Thus, the probability distribution $p^{S,\Delta}(k,l)$ will be a sum of the distributions obtained for $S_{i_1}$ and $S_{i_2}$ separately, see Fig.~\ref{line}b. The filter projects each Fock state coherently on a line, but projections on two different lines are incoherent with respect to each other
\begin{align}
  \label{P_mixture}
  &\mathcal{P}^{\{S_j,\Delta_j\}_{j=1,2}}_{\text{C}}[\{S_{i_{j}},\Delta_{i_{j}}\}_{j=1,2}] \\
  &\qquad= q \mathcal{P}^{S_1,\Delta_1}_{\text{C}}[S_{i_1},\Delta_{i_1}] + (1-q) \mathcal{P}^{S_2,\Delta_2}_{\text{C}}[S_{i_2},\Delta_{i_2}].\nonumber
\end{align}
  
Let us now consider a superposition state where each term has fixed photon number $S_i$ and population difference is distributed uniformly
\begin{equation}
\label{psi1}
|\psi_i\rangle = \tfrac{1}{\sqrt{S_i+1}}\sum_{N=0}^{S_i} |S_i-N, N\rangle.
\end{equation}
Since the photon number $S_i$ is known, in this case the nonzero elements of $p^{S,\Delta}(k,l)$ must be located on a line as well.  However, since the terms with negative and positive $\Delta_i$ contribute to this state equally, the probability distribution is symmetric with respect to the line $\Delta_t=0$. Two generic examples of such distributions are depicted in Fig.~\ref{line}c \& d. The distribution shown in Fig.~\ref{line}c is obtained if behind the PBS, the measured population difference roughly equals the sum of the reflected photons ($|\Delta|\simeq 0$), whereas the distribution from Fig.~\ref{line}d applies if $|\Delta|\simeq S$. Again, the projection of the initial superposition on the line $l=S_t-k$, where $S_t=S_i-S$, performed by the filter is coherent. The action of the setup in Fig.~\ref{filter} on a superposition with a fixed photon number and distribution of population difference is described by following  sum of the operators (\ref{P_Fock}) 
\begin{align}
\mathcal{P}^{S,\Delta}_{\text{C}}[S_i] = \sum_{\Delta_i=-S_i}^{S_i} \mathcal{P}^{S,\Delta}_{\text{C}}[S_i,\Delta_i].
\end{align}

\begin{figure}
\begin{center}
  \includegraphics[width=3.5cm]{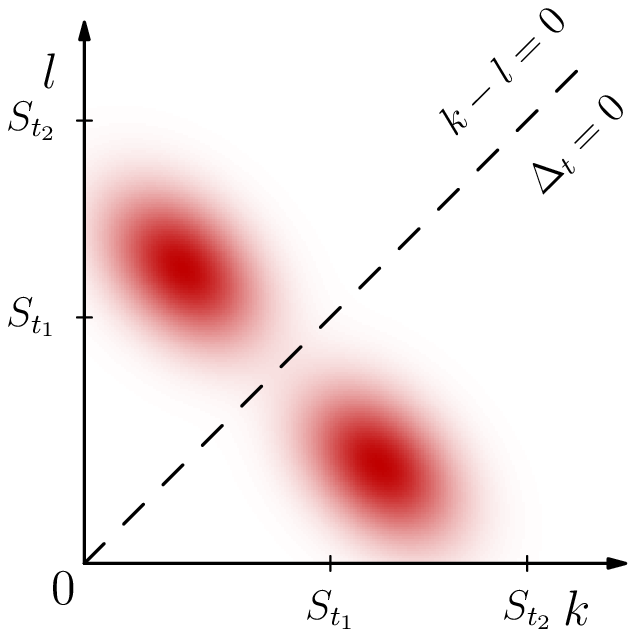}\quad
  \includegraphics[width=3.5cm]{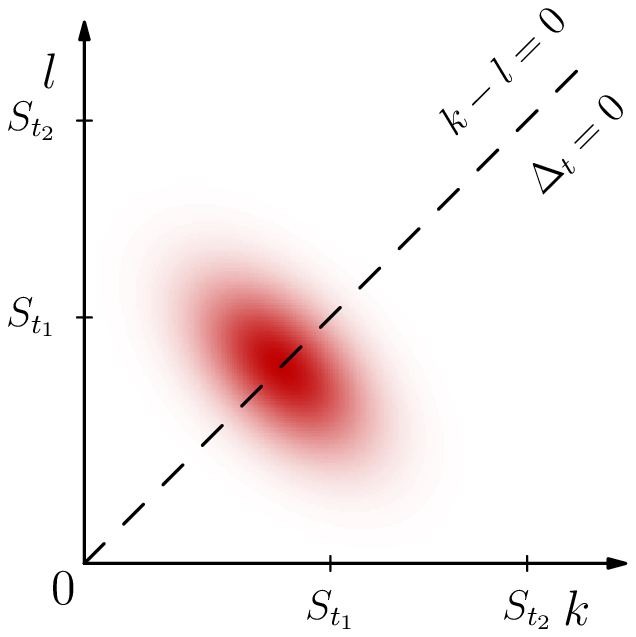}
\end{center}
\caption{Visualization of the photon number distribution for the transmitted beam \emph{before} the shutter in Fig.~\ref{filter}b for an input state with a uniform distribution of the initial population difference and the total photon number $\lvert\psi_i\rangle=\tfrac{1}{\sqrt{S_{i_2}-S_{i_1}+1}}\sum_{S_i=S_{i_1}}^{S_{i_2}}\tfrac{1}{\sqrt{S_i+1}} \sum_{N=0}^{S_i} \lvert S_i-N, N\rangle$. After the projection the coherence is preserved within the colored areas.}
\label{area}
\end{figure}

Another important example is a state with a uniform distribution of both, the initial population difference and the total photon number
\begin{equation}
\label{psi2}
\lvert\psi_i\rangle=\tfrac{1}{\sqrt{S_{i_2}-S_{i_1}+1}}\sum_{S_i=S_{i_1}}^{S_{i_2}}\tfrac{1}{\sqrt{S_i+1}} \sum_{N=0}^{S_i} \lvert S_i-N, N\rangle.
\end{equation}
The uniform distribution is the worst case scenario with respect to the amount of information it carries about the variable.  Now from Eq.~(\ref{wz6}) we see that $|\psi_t\rangle$ is a superposition of the following terms $|S_i-N-v, N-w\rangle$, where $S_t=S_i-S$ and $\Delta_t = \Delta_i + w - v$. We note that $S_t\in[S_{i_1}-S, S_{i_2}-S]$ and $\Delta_t \in [-S_t, S_t]$. Thus, there is more than one $S_i$ (and $S_t$) which contributes to the same $\Delta_t$. While computing photon number distribution for $|\psi_t\rangle$ we take projection on such terms simultaneously, i.e. we add their amplitudes of probability. Therefore, now the projections on different lines of photon number $S_t$ are \emph{coherent}: the filter projects onto a certain area in space $(k,l)$ or $(S_t,\Delta_t)$, see Fig.~\ref{area}. We conclude that the quantum filter performs the following operation on a general two-mode input states 
\begin{align}
  \label{P_uniform}
\mathcal{P}_{\text{C}}^{S,\Delta} = \sum_{S_i=0}^{\infty}\sum_{\Delta_i=-S_i}^{S_i} \mathcal{P}^{S,\Delta}_{\text{C}}[S_i,\Delta_i].
\end{align} 
This operator plays the role of the Kraus operator for the filter.

Please note that regardless the input state $\lvert\psi_i\rangle =\sum_{S_i,\Delta_i}\xi_{S_i, \Delta_i}\lvert S_i, \Delta_i\rangle$, the setup in Fig.~\ref{filter} preserves the symmetry of the initial state in the photon number space. This follows from the fact that the incoming, the reflected and the transmitted beams are correlated. The filter convolutes the initial photon number statistics with the beam splitter probability distribution which is symmetric with respect to the population difference $\xi_{S_i, \Delta_i} \to \xi_{S_i, \Delta_i}\cdot f^{S,\Delta}_{S_i,\Delta_i}(\Delta_r=\Delta_i-\Delta_t)$, (see Fig.~\ref{fig:single_mzi_s200}). This results in a ``blurred'' photon number statistics of the transmitted beam with respect to the initial one
\begin{align}
 \lvert\psi_t\rangle &{}= \mathcal{P}^{S,\Delta}_{\text{C}} \lvert\psi_i\rangle \label{blee}\\
&{}= \sum_{S_i,\Delta_i} \xi_{S_i, \Delta_i} \lvert S_i-S\rangle \nonumber\\
&\otimes \left( \sum_{\substack{\Delta_r=\Delta_r^{min}\\\text{C}(|\Delta_r|)}}^{\Delta_r^{max}} f^{S,\Delta}_{S_i,\Delta_i}(\Delta_r) 
\lvert \Delta_i-\Delta_r\rangle\right).\nonumber
\end{align}

\begin{figure}
\begin{center}
  \raisebox{3.6cm}{a)}\includegraphics[width=3.5cm]{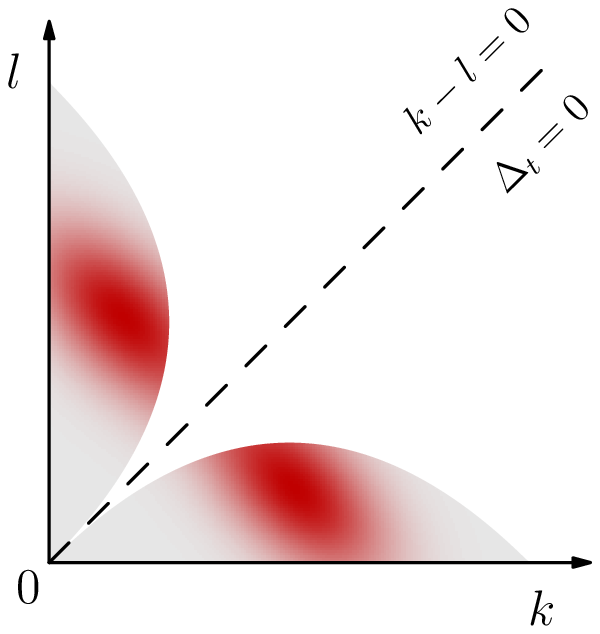}\quad
  \raisebox{3.6cm}{b)}\includegraphics[width=3.5cm]{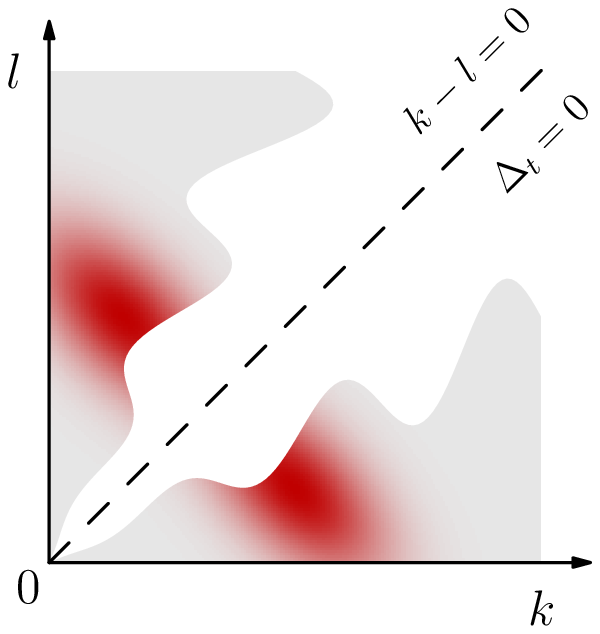}\quad
\\
  \raisebox{3.6cm}{c)}\includegraphics[width=3.5cm]{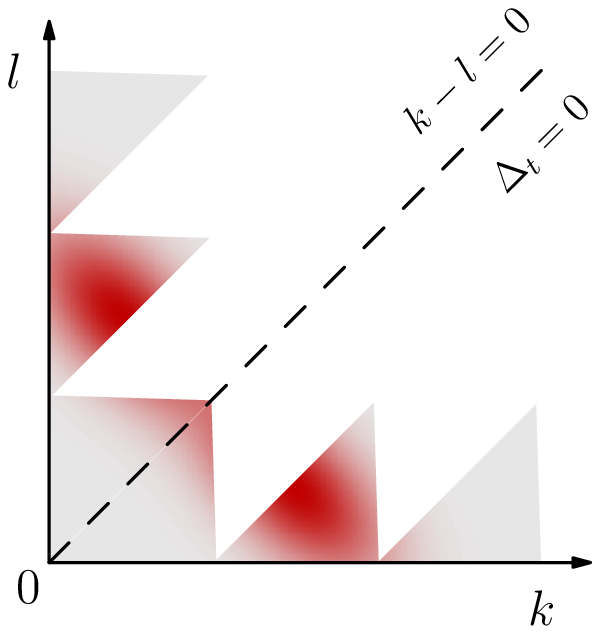}\quad
  \raisebox{3.6cm}{d)}\includegraphics[width=3.5cm]{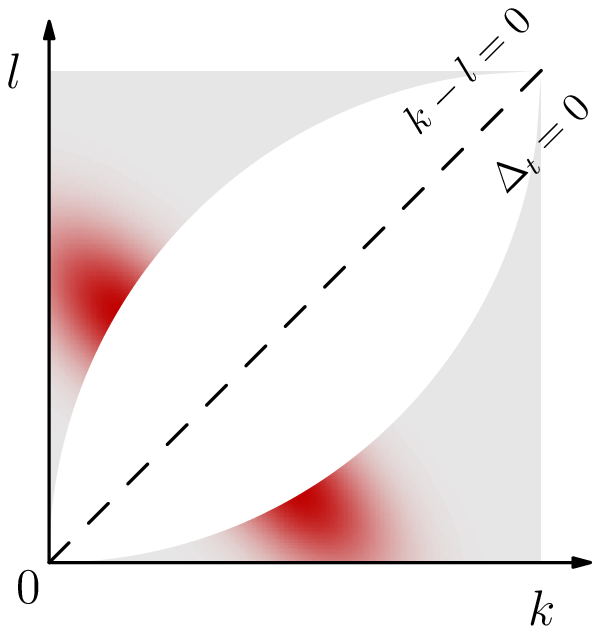}
\end{center}
\caption{Exemplary projection areas selected with the quantum filter presented on Fig.~\ref{filter} using filtering condition $\text{C}(S_t,|\Delta_t|)$ applied by the analysis box and the shutter: a) $|\Delta_t| > (aS_t)^2$, b) $|\Delta_t| > aS_t(1 + b\sin(cS_t))$, c) $|\Delta_t| > \lfloor aS_t\rfloor/a$, d) $|\Delta_t| > S_t + \sqrt{b^2 - S_t^2} - b$.  Here, $0 < a < 1$ and $b>0$ are real parameters.}
\label{filters}
\end{figure}

We would like to comment on the filtering condition $\text{C}(|\Delta_r|)$ present in the operators in Eqs.~(\ref{P_Fock})- (\ref{blee}).  It directly results from the condition $\text{C}(S_t,|\Delta_t|)$ ($\Delta_t=\Delta_i-\Delta_r$). So far, the influence of this condition was not discussed. It is an additional handle which allows to shape the areas and line of projection shown in Figs.~\ref{line}-\ref{area}. The physical implementation of the filter in Fig.~\ref{filter} allows to impose an arbitrary filtering condition $C$, symmetric with respect to the line $k=l$ ($\Delta_t=0$), on the sum $S_t=k+l$ and the modulus of the difference $|\Delta_t|=|k-l|$.  The fact that $C$ is symmetric results from the Hong-Ou-Mandel interference. Fig.~\ref{filters} depicts exemplary projection areas for various filtering conditions.

\section{Quantum filtering in presence of imperfect photodetection}
\label{photodetection}

The analysis of the operation performed by the family of non-Gaussian filters on the input quantum state presented in Section \ref{sec:efficient_filtering} has not taken into account any imperfections in the measurement process. In Appendix~D of \cite{MDF} we considered a simple case of losses in the system, modeled with an additional beam splitter put before the shutter. However, that model still assumed the precise measurement of $S$ and $\Delta$ by the perfect photon counting detectors.  These parameters, as shown above, are of a great importance in the process of finding an output photon number distribution of the transmitted beam $p^{S,\Delta}(S_t,\Delta_t)$.  Therefore, in this section we will discuss the influence of the imperfect photodetection on the filter output.

Detection imperfections in the setup presented in Fig.~\ref{filter}b could be caused for example by the errors in the photon counting process (e.g.\ arising in the detector electronics). They result in lower (e.g. losses) or greater (e.g. dark counts) number of registered photons than expected.  This process is independent for both detectors placed at the outputs of the polarizing beam splitter (Fig.~\ref{filter}a).  As a result, a detector could register a different Fock state $\lvert K'\rangle$ than $\lvert K\rangle$ which really leaved the PBS. This is represented by some distribution $d_K(K')$ giving the probability of registering the state $\lvert K'\rangle$ instead of state $\lvert K\rangle$.  Here, $\sum_{K'} d_K(K')=1$.  The distribution $d_K(K')$ represents a detector characteristics.  It transforms the quantum state before the shutter $\lvert\psi_t\rangle$ (Eq.~(\ref{bzyk})) to a mixed state $\rho'_t$
\begin{align}
\rho'_t &= \sum_{K', L'} d_K(K')\,d_L(L') \big|{}_d\langle K',L'\vert\psi_{dt}\rangle\big|^2.
\label{mixedfiltered}
\end{align}
The imperfect detectors, which turn the pure output quantum state into a mixed one, may significantly affect the coherent action of the filter. According to the formula (\ref{mixedfiltered}), the filter will now perform the following operation 
\begin{align}
  \label{P_uniform_mixed}
\mathcal{P}_{\text{C}}&{}= \sum_{S', \Delta'} d_{S,\Delta}(S',\Delta') \mathcal{P}_{\text{C}}^{S',\Delta'},
\end{align} 
with $d_{S,\Delta}(S',\Delta') = d_{\tfrac{S-\Delta}{2}}(\tfrac{S'-\Delta'}{2})d_{\tfrac{S+\Delta}{2}}(\tfrac{S'+\Delta'}{2})$.
In order to examine its influence on the output state, it is necessary to compute the purity of $\rho'_t$: $\gamma=\mathrm{Tr}\{(\rho'_t)^2\}$.

In our first model, the detector non-unit efficiency $\eta<1$ is simulated with an additional beam splitter put in front of each ideal photon counting detector.  The transmitivity of these beam splitters is equal to $\eta$.  The detector, instead of projecting the incoming beam on the Fock state $\lvert K\rangle_d$, projects on a mixture of Fock states
\begin{align}
  |K\rangle\langle K|&\to\mathrm{Tr}_{loss}\{\mathcal{U}_{BS}\lvert K, 0\rangle\}\\
&= \sum_{x=0}^K \binom{K}{x} (1-\eta)^{x} \eta^{K-x} \lvert K-x\rangle\langle K - x|,\nonumber
\end{align}
with the binomial distribution
\begin{equation}
\label{binomial_distr}
d_K(K')=\binom{K}{K'} (1-\eta)^{K-K'} \eta^{K'}.
\end{equation}
In the limit of $\eta\to1$, $d_K(K')\to\delta_{K,K'}$ -- the Kronecker delta, which gives the result for perfect detectors. Similar results are obtained for detector measuring $|L\rangle_d$. Please note that this photodetection model assumes that $K$ ($L$) is the maximal possible measurement result. The most probable result is $\eta\cdot K$. Therefore, this model includes only losses in the photodetector.

The second model of the imperfect detector is described by a Gaussian distribution of a given standard deviation $\sigma$. This corresponds to the Weierstrass transform (known as the Gaussian blur) applied to the photon number distribution measured by the ideal detectors
\begin{equation}
\label{gaussian_distr}
    d_K(K') = \dfrac{1}{\sqrt{2\pi\sigma^2}} e^{\tfrac{-(K-K')^2}{2\sigma^2}}.
\end{equation}
In the limit of $\sigma\to 0$, $d_K(K')\to\delta(K-K')$ -- the Dirac delta, which corresponds to the perfect detection.  This model assumes that the most probably event is the detection of the actual photon number $K$. However, it takes into account that the detector, with equal probabilities, can measure higher and lower photon numbers. The detection of higher photon number $K'>K$ may happen due to dark counts or cross-talks in the separate channels of photodetectors.

\section{Numerical results}
\label{sec:results}

In order to illustrate the action of the family of non-Gaussian quantum filters, we performed numerical computations for two input quantum states. Below, we will present the photon number distributions for the output states \emph{before} the shutter, which result from two important examples of quantum input states discussed in Section III. The first state is a uniform superposition of Fock states of a constant photon number $S_i$, given by Eq.~(\ref{psi1}). The second analyzed state is a uniform superposition of states (\ref{psi1}) with the photon number between $S_{i_1}$ and$ S_{i_2}$, given by Eq.~(\ref{psi2}).

\begin{figure}[H]
\begin{center}
  \raisebox{3cm}{a)}\includegraphics[width=3.8cm]{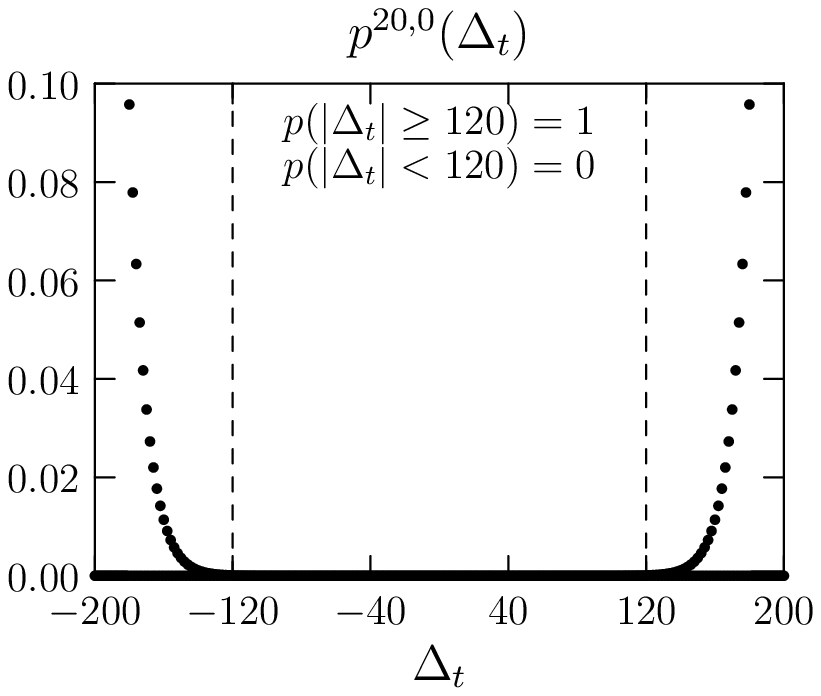}
  \raisebox{3cm}{b)}\includegraphics[width=3.8cm]{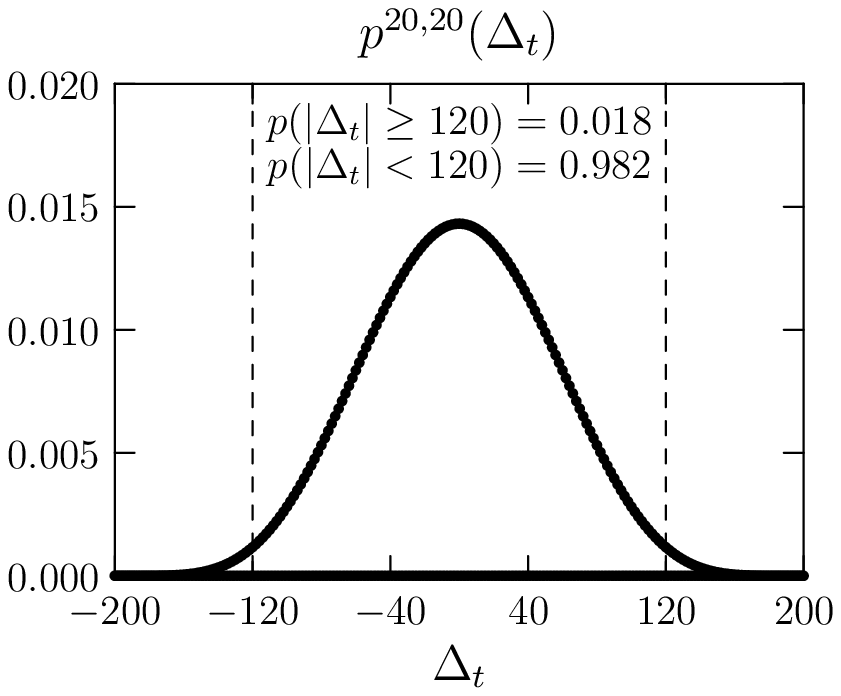}
\\
  \raisebox{3cm}{c)}\includegraphics[width=3.8cm]{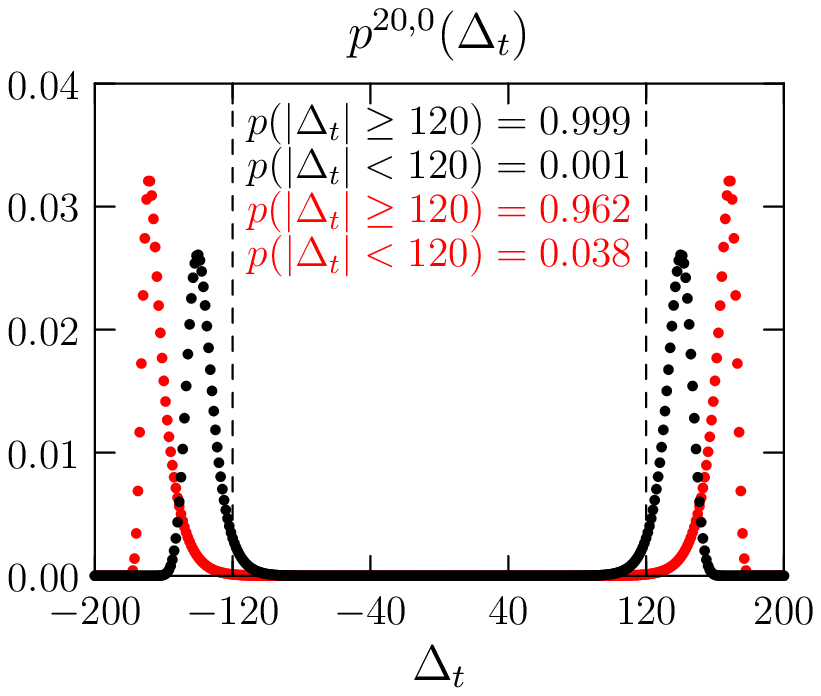}
  \raisebox{3cm}{d)}\includegraphics[width=3.8cm]{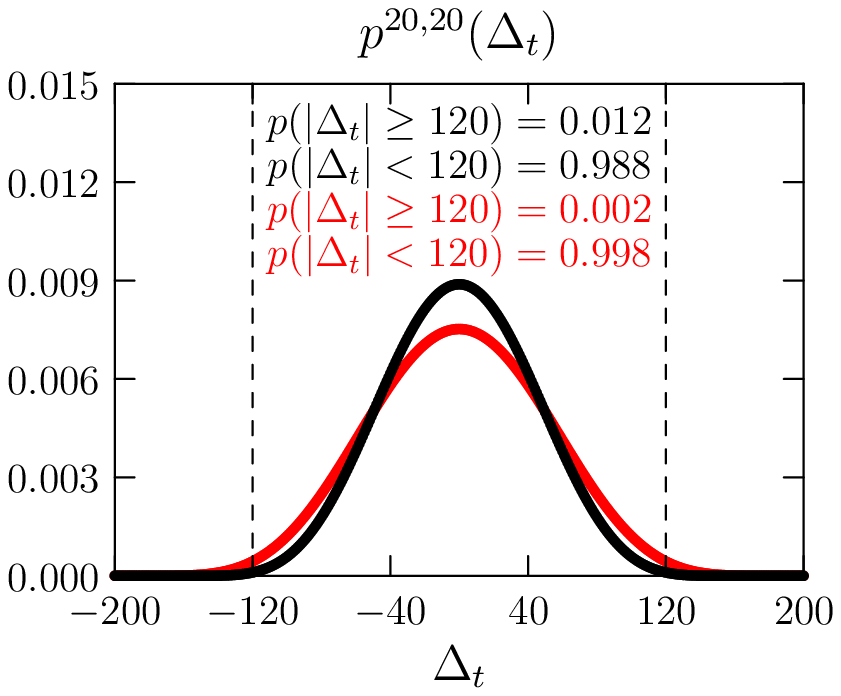}
\\
  \raisebox{3cm}{e)}\includegraphics[width=3.8cm]{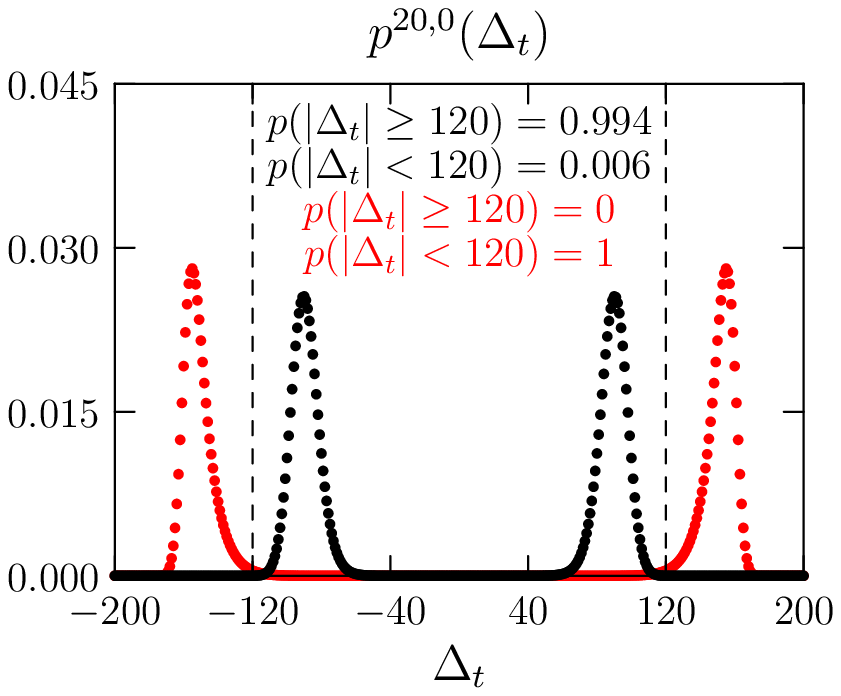}
  \raisebox{3cm}{f)}\includegraphics[width=3.8cm]{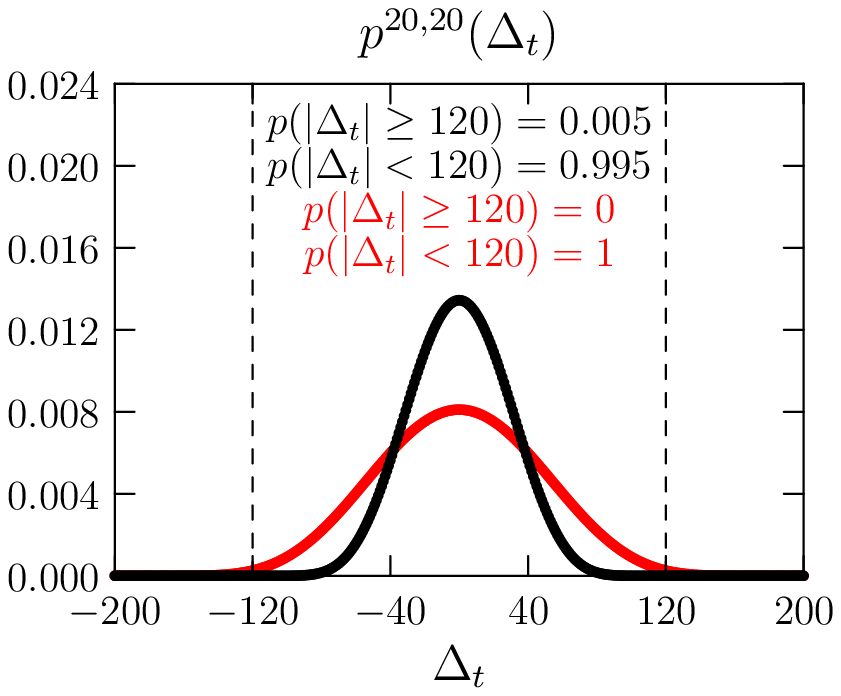}
\end{center}
\caption{The plots of probability distributions $p^{S,\Delta}(\Delta_t)$ numerically computed for the input state (\ref{psi1}) before the shutter, with a constant total number of $S_i=200$, $r=10\%$, $S=20$, $\Delta=0$ (left column) and $\Delta=20$ (right column). The results were obtained for both perfect (a \& b) and imperfect photodetection, with binomial (c \& d) and Gaussian (e \& f) distribution $d_K(K')$ ($d_L(L')$), representing the detector characteristics. Black and red curves were obtained for different values of parameters $\eta$ and $\sigma$. Detailed description is presented in the main text.}
\label{psi1_Si200}
\end{figure}

Fig.~\ref{psi1_Si200} depicts the plots of probability distributions $p^{S,\Delta}(\Delta_t)$ computed for the input state (\ref{psi1}) before the shutter, with a constant total number of $S_i=200$ photons, reflectivity of a tapping beam splitter $10\%$ and $S=20$ photons registered at the detectors. Left column (plots a), c) and e)) contains the distributions obtained for $\Delta=0$, whereas right column (plots b), d) and f)) -- $\Delta=S=20$. Plots a) \& b) show the distributions in case of the ideal detectors; plots c) \& d) -- lossy detectors with efficiency $\eta=95\%$ (red) and $\eta=80\%$ (black); plots e) \& f) -- imprecise detectors with the Gaussian distribution of the standard deviation $3\sigma=5$ (red) and $3\sigma=20$ (black).

The probability distributions computed for the above cases allow to predict the probability of meeting a given condition $\text{C}(S_t,|\Delta_t|)$. This condition can by arbitrary and chosen in order to optimize performance of a certain quantum task. For example, let us assume that in some quantum application we need a quantum state with difference of population between the modes larger than 120 photons, i.e. $\text{C}(S_t,|\Delta_t|)\equiv \{|\Delta_t|\geq 120\}$. States filtered according to a condition that the population difference between its two modes is greater than some threshold value are realization of superpositions of the N00N and NmmN states, which find applications in quantum metrology for enhanced optical phase estimation~\cite{Demkowicz}. If we know that the source produces a superposition of uniformly distributed Fock states (Eq.~(\ref{psi1})) of total number of $S_i=200$ photons, the measurement result of $S$ and $\Delta$ gives us the information of the probability of fulfilling $\text{C}(S_t,|\Delta_t|)$.  Here, if detectors were perfect, measurement of $S=20$, $\Delta=0$ would give us a certainty that the condition is met (Fig.~\ref{psi1_Si200}a), whereas $\Delta=20$ would inform that is not fulfilled with the probability of $0.982$ (Fig.~\ref{psi1_Si200}b). Therefore, when $\Delta=0$ the box should open the shutter and close it when $\Delta=20$.  Similar analysis would be performed for all possible values of $S$ and $\Delta$.

However, imperfections in the detectors influence the results. In case of detector efficiency $\eta$ modeled by the binomial distribution given by Eq.~(\ref{binomial_distr}), the probability of $|\Delta_t|\geq 120$ conditioned on $\Delta=0$ lowers to $0.999$ for $\eta=5\%$ and $0.962$ for $\eta=20\%$ (Fig.~\ref{psi1_Si200}c). In the same time the probability of not fulfilling $\text{C}(S_t,|\Delta_t|)$ when $\Delta=20$ raises to $0.988$ for $\eta=5\%$ and even $0.998$ for $\eta=20\%$ (Fig.~\ref{psi1_Si200}d).  Similar results are obtained in case of the imperfections modeled with the Gaussian distribution (Eq.~(\ref{gaussian_distr})), but here $p(|\Delta_t|\geq 120)=0.994$ for $\Delta=0$, $3\sigma=5$ and $0.962$ for $3\sigma=20$ (Fig.~\ref{psi1_Si200}e).  Finally, Fig.~\ref{psi1_Si200}f says that observing $\Delta=20$ gives us probability $0.995$ that $|\Delta_t|$ for $3\sigma=5$ and $1$ for $3\sigma=20$.

\begin{figure}[p]
\begin{center}
  \raisebox{3cm}{a)}\includegraphics[width=3.8cm]{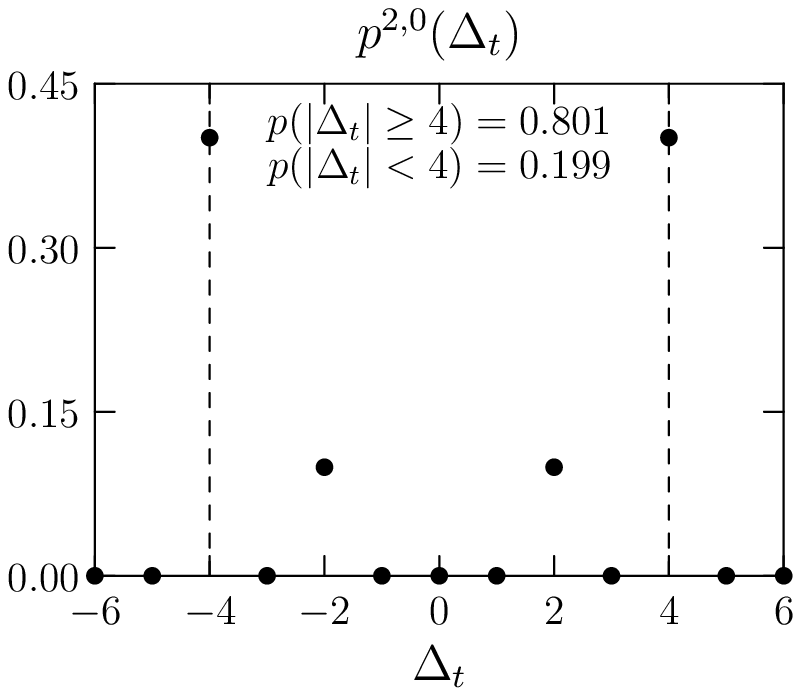}
  \raisebox{3cm}{b)}\includegraphics[width=3.8cm]{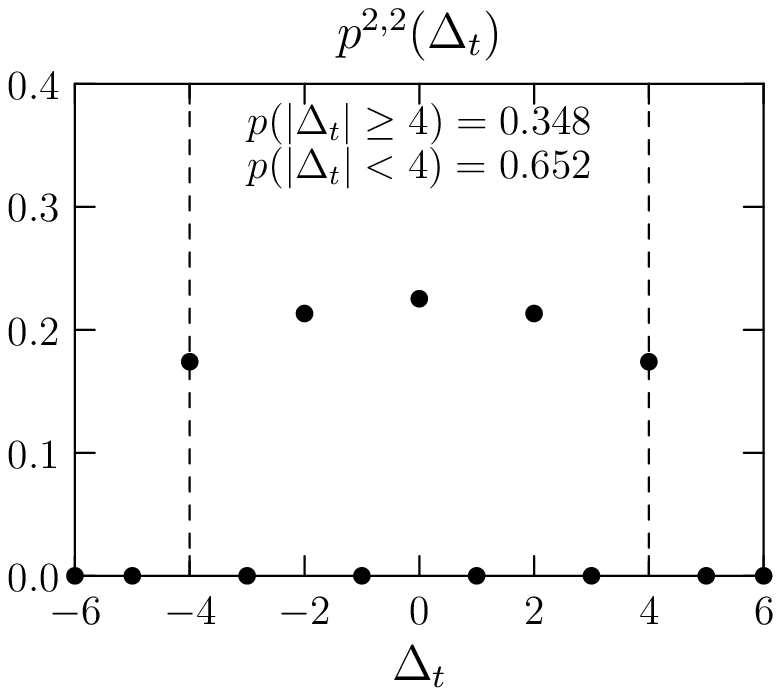}
\\
  \raisebox{3cm}{c)}\includegraphics[width=3.8cm]{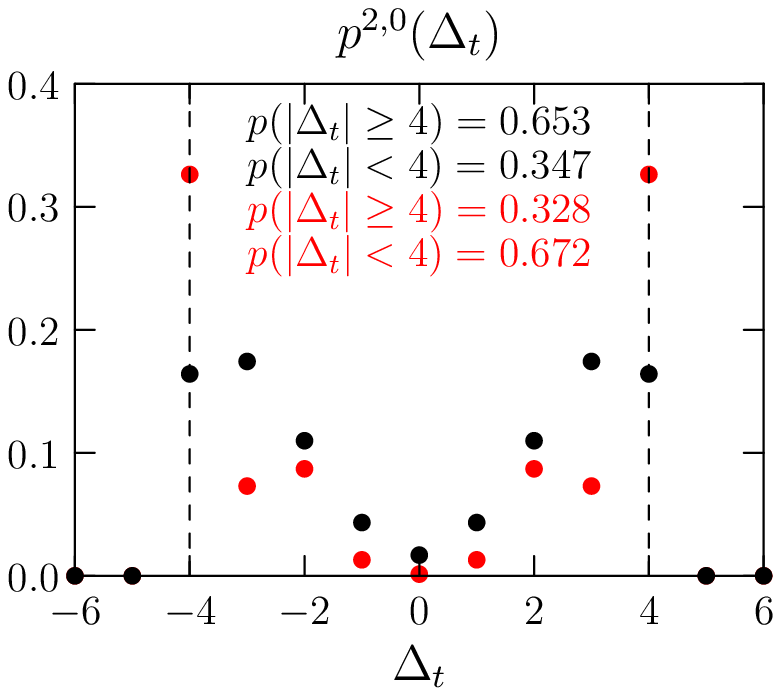}
  \raisebox{3cm}{d)}\includegraphics[width=3.8cm]{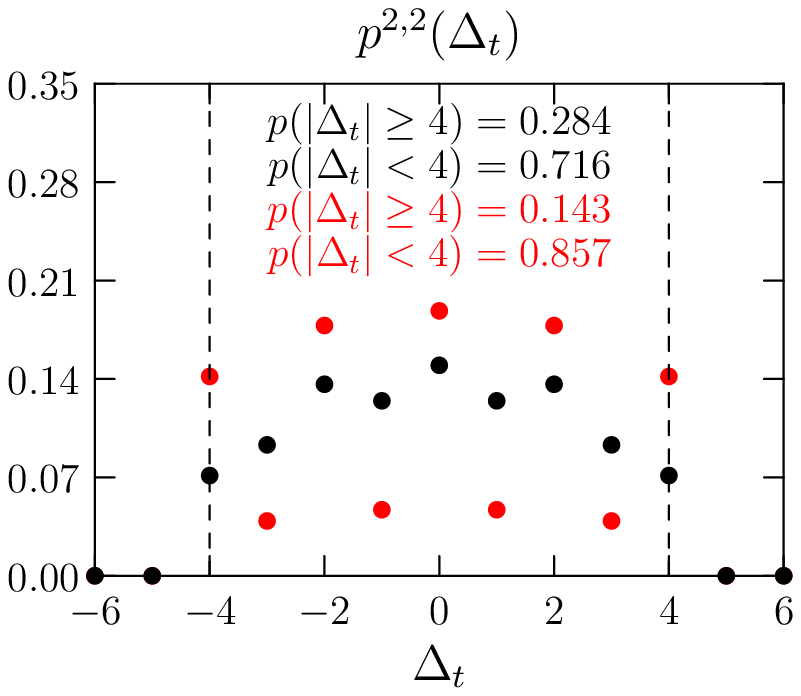}
\\
  \raisebox{3cm}{e)}\includegraphics[width=3.8cm]{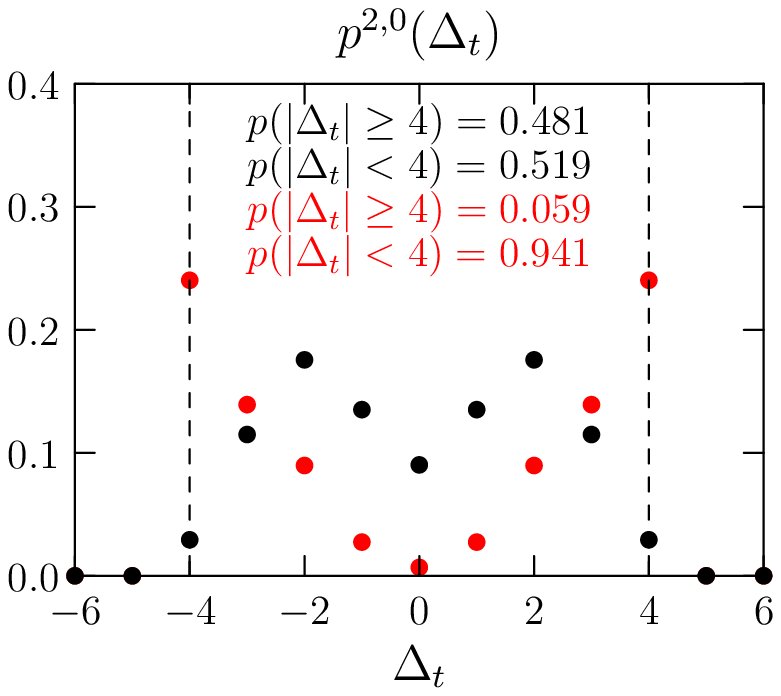}
  \raisebox{3cm}{f)}\includegraphics[width=3.8cm]{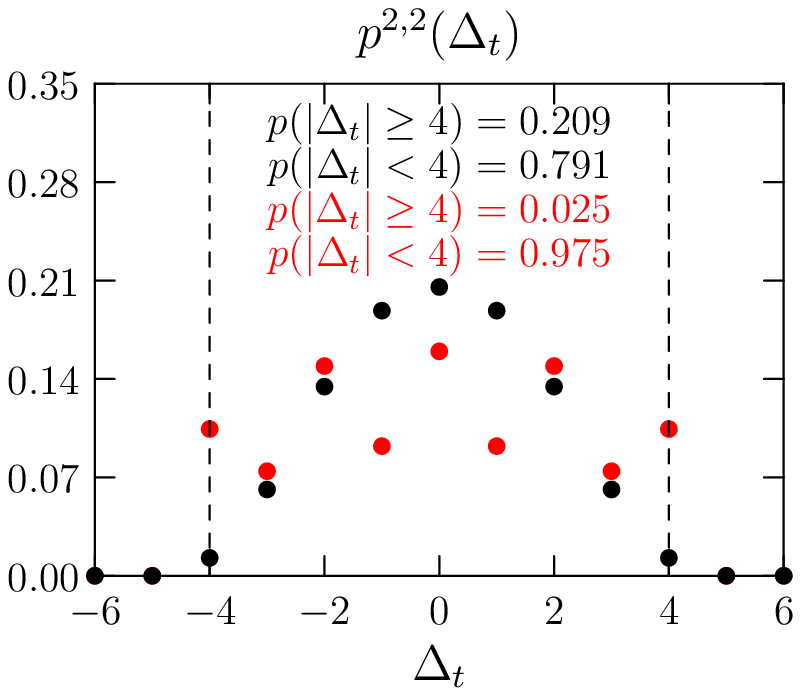}
\end{center}
\caption{The plots of probability distributions $p^{S,\Delta}(\Delta_t)$ numerically computed for the input state (\ref{psi1}) with a constant total number of $S_i=6$, $r=10\%$, $S=2$, $\Delta=0$ (left column) and $\Delta=2$ (right column). The results were obtained for both perfect (a \& b) and imperfect photodetection, with binomial (c \& d) and Gaussian (e \& f) distribution $d_K(K')$ ($d_L(L')$), representing the detector characteristics. Black and red curves were obtained for different values of parameters $\eta$ and $\sigma$. Detailed description is presented in the main text.}
\label{psi1_Si6}
\end{figure}

Fig.~\ref{psi1_Si6} depicts the same collection of probability distributions $p^{S,\Delta}(\Delta_t)$ of the state (\ref{psi1}) before the shutter, but computed for a small photon number $S_i=6$.  Reflectivity of a tapping beam splitter remainded $10\%$ and we assumed $S=2$ photons registered at the detectors. Left column (plots a), c) and e)) contains the distributions obtained for $\Delta=0$, whereas right column (plots b), d) and f)) -- $\Delta=S=2$. Plots a) \& b) show the distributions in case of ideal detectors; plots c) \& d) -- lossy detectors with efficiency $\eta=95\%$ (red) and $\eta=80\%$ (black); plots e) \& f) -- imprecise detectors with the Gaussian distribution of the standard deviation $3\sigma=0.15$ (red) and $3\sigma=0.6$ (black).  Here we computed the probabilities of events that $|\Delta_t|$ exceeds or is below threshold equal to $4$.  

\begin{figure}[p]
\begin{center}
  \raisebox{4cm}{a)}\includegraphics[width=3.8cm]{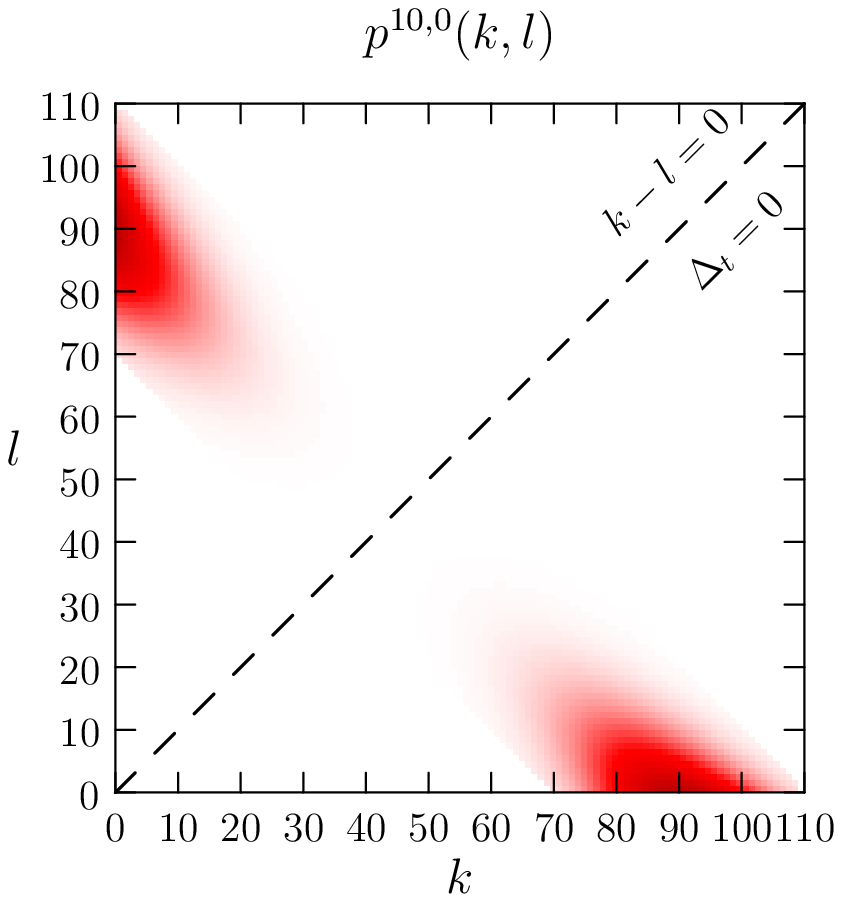}
  \raisebox{4cm}{b)}\includegraphics[width=3.8cm]{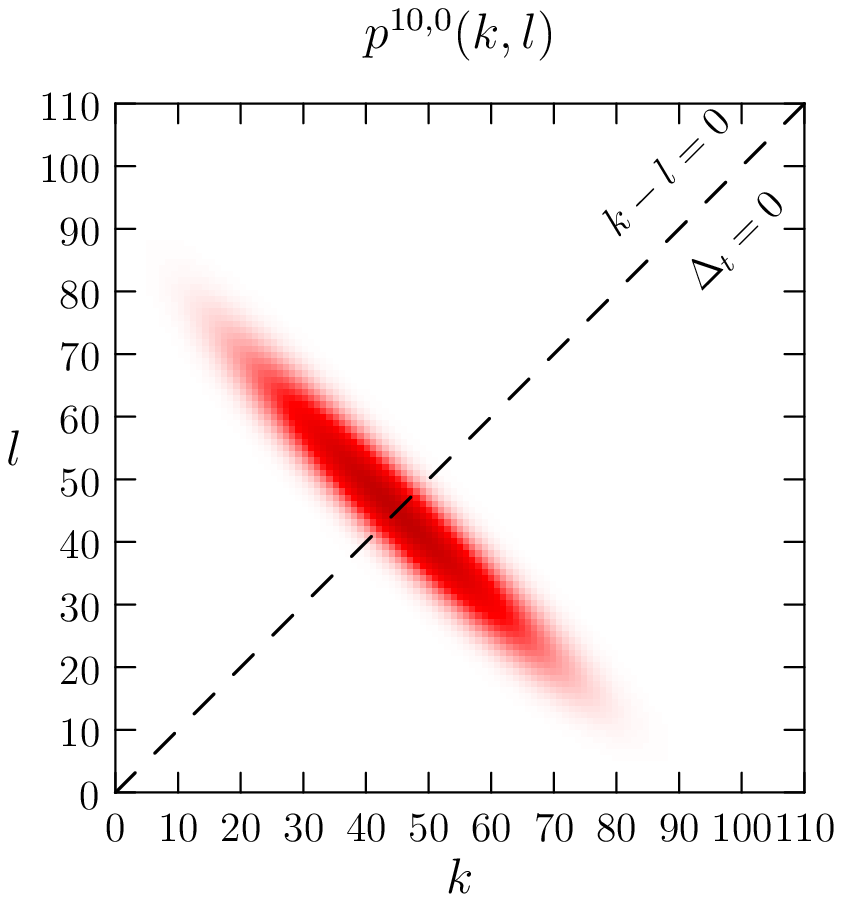}
\\
  \raisebox{4cm}{c)}\includegraphics[width=3.8cm]{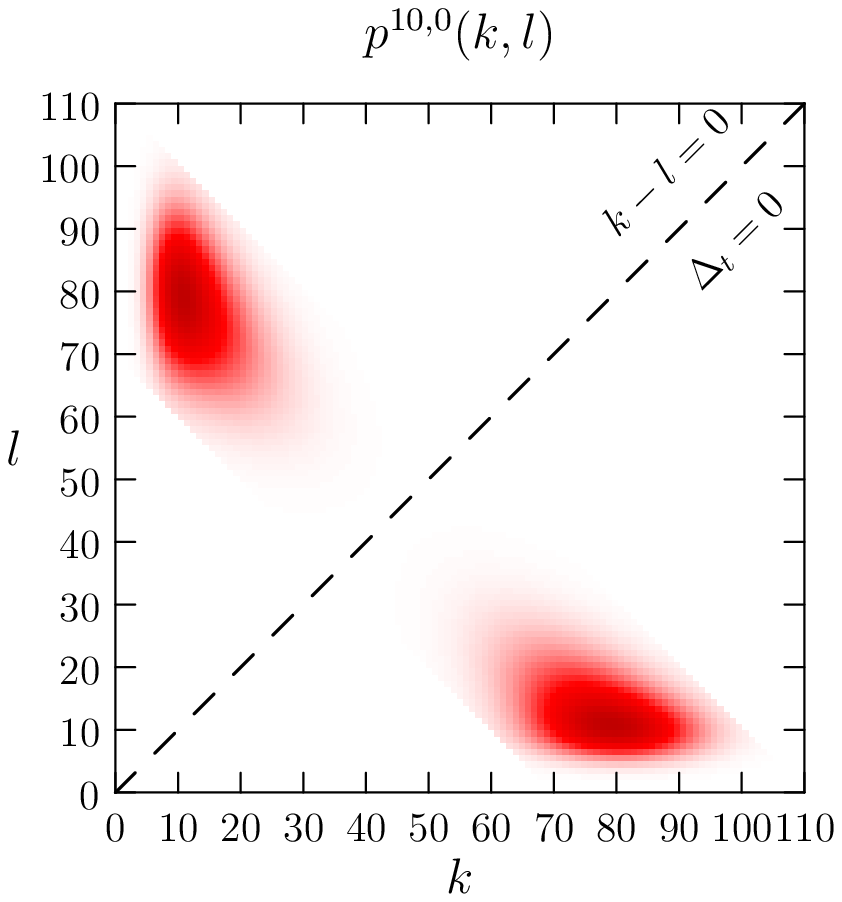}
  \raisebox{4cm}{d)}\includegraphics[width=3.8cm]{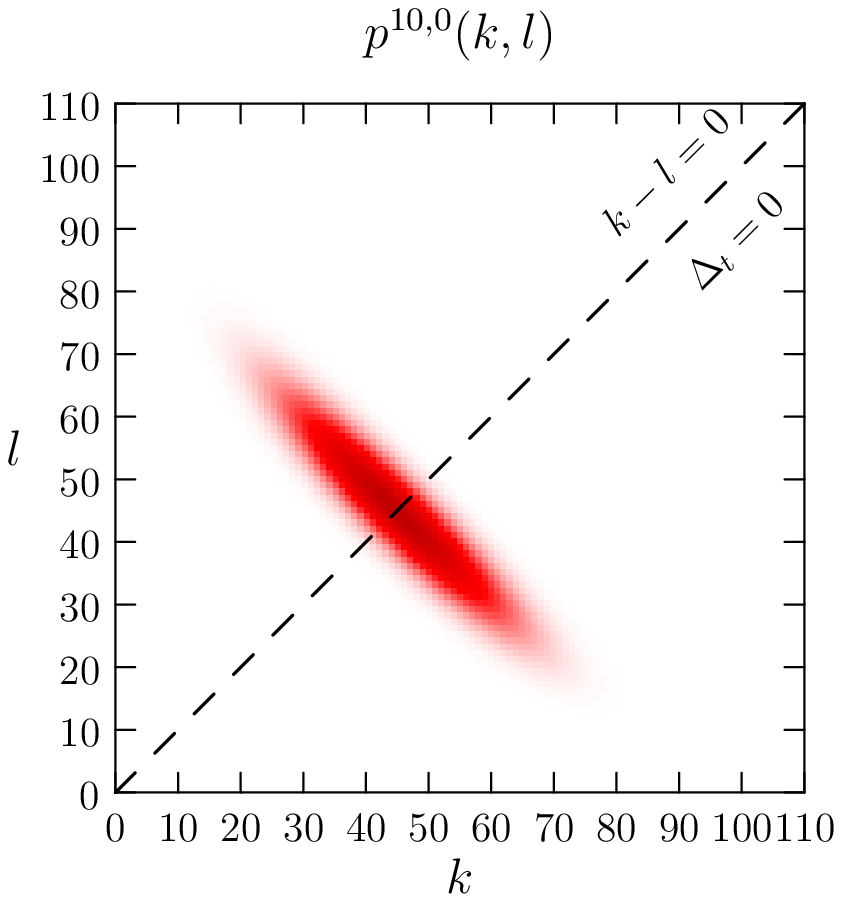}
\\
  \raisebox{4cm}{e)}\includegraphics[width=3.8cm]{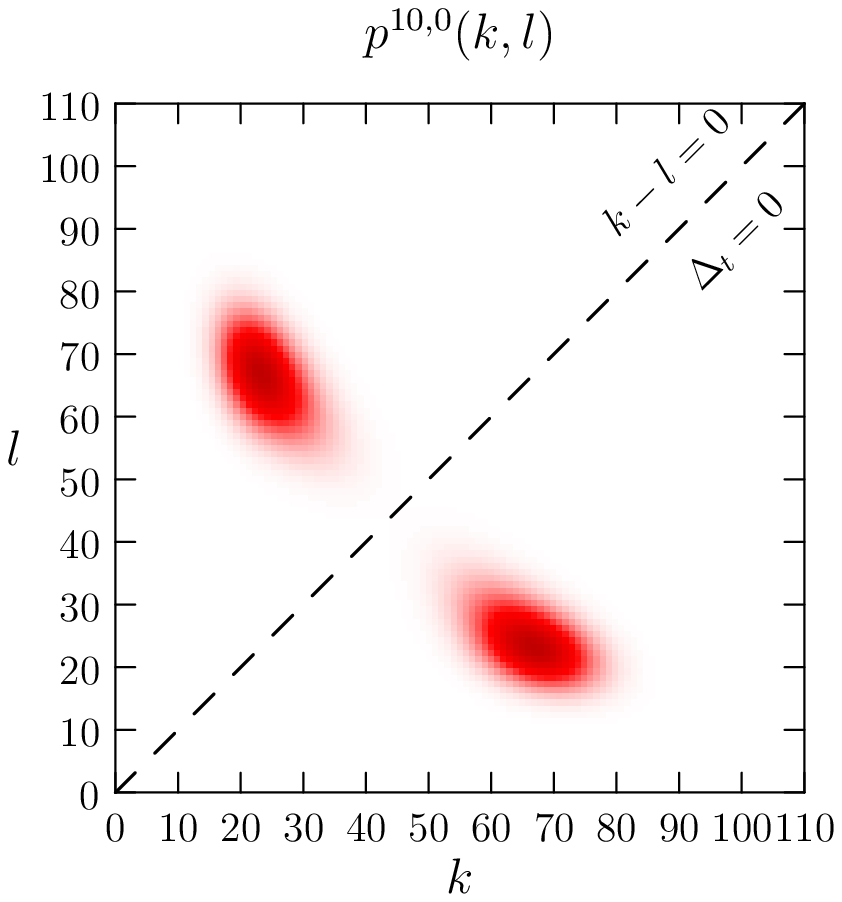}
  \raisebox{4cm}{f)}\includegraphics[width=3.8cm]{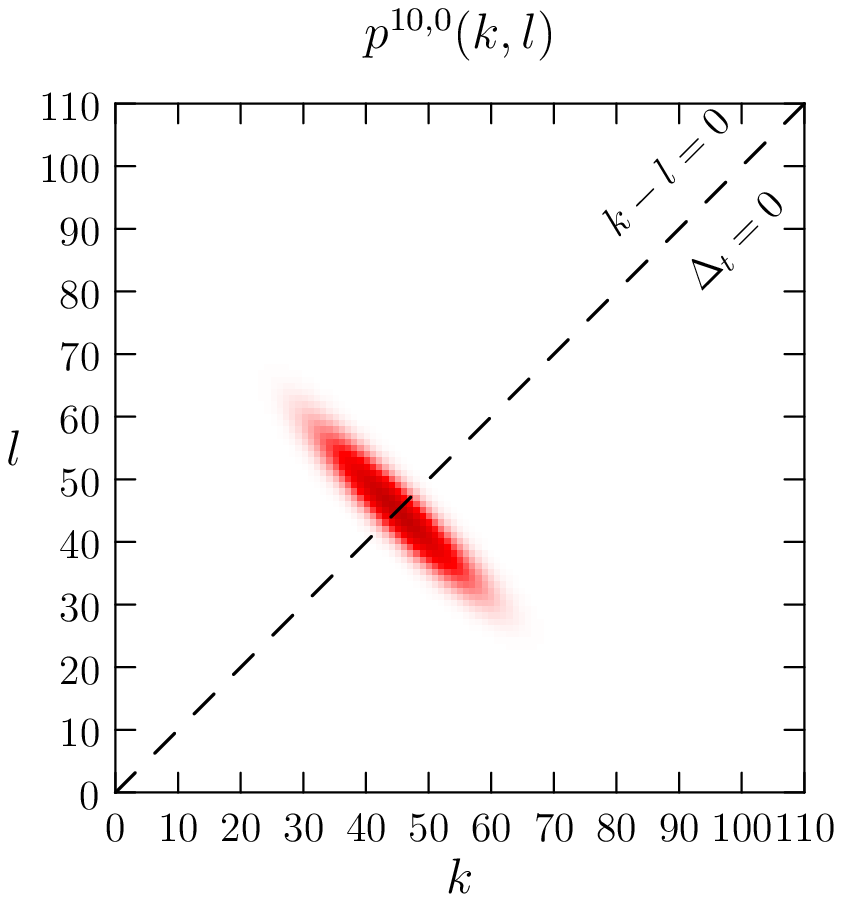}
\end{center}
\caption{The plots of probability distributions $p^{S,\Delta}(\Delta_t)$ numerically computed for the input state (\ref{psi2}) with a total number of photons in range $S_i\in[80,120]$, $r=10\%$, $S=10$, $\Delta=0$ (left column) and $\Delta=10$ (right column). The results were obtained for both perfect (a \& b) and imperfect photodetection, with binomial (c \& d) and Gaussian (e \& f) distribution $d_K(K')$ ($d_L(L')$), representing the detector characteristics. Detailed description is presented in the main text.}
\label{psi2_Si80_120}
\end{figure}

\begin{figure}[p]
\begin{center}
  \raisebox{4cm}{a)}\includegraphics[width=3.8cm]{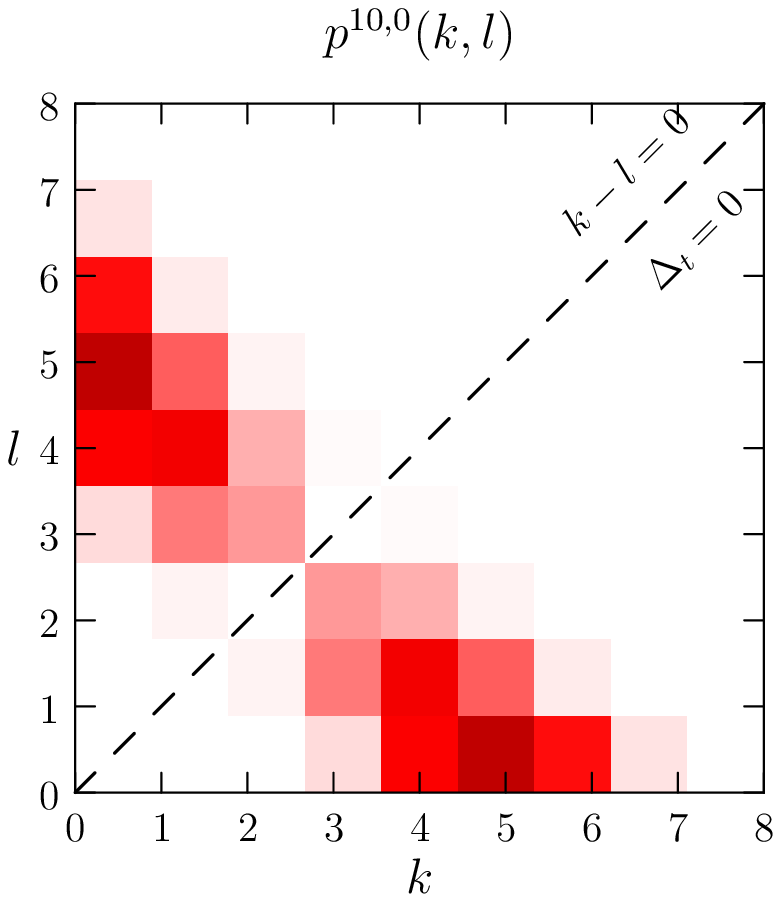}
  \raisebox{4cm}{b)}\includegraphics[width=3.8cm]{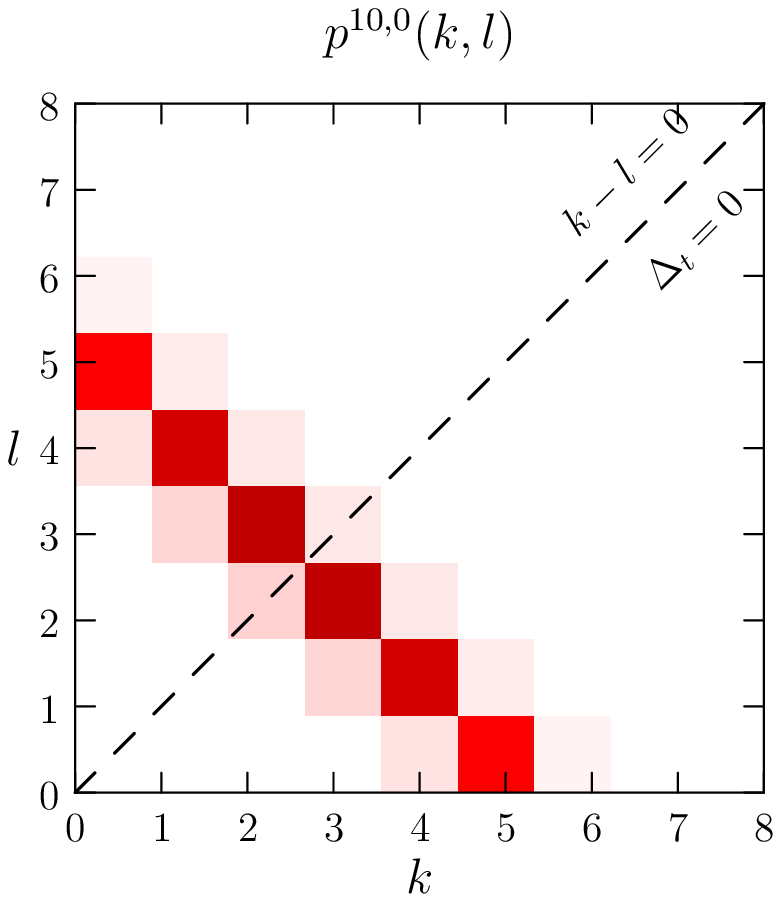}
\\
  \raisebox{4cm}{c)}\includegraphics[width=3.8cm]{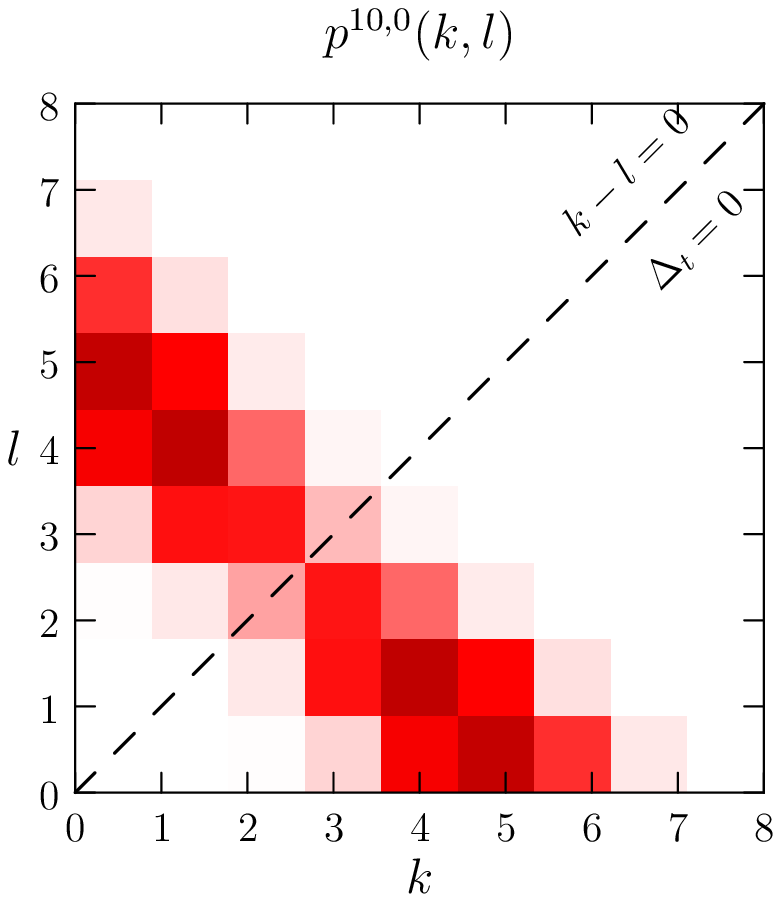}
  \raisebox{4cm}{d)}\includegraphics[width=3.8cm]{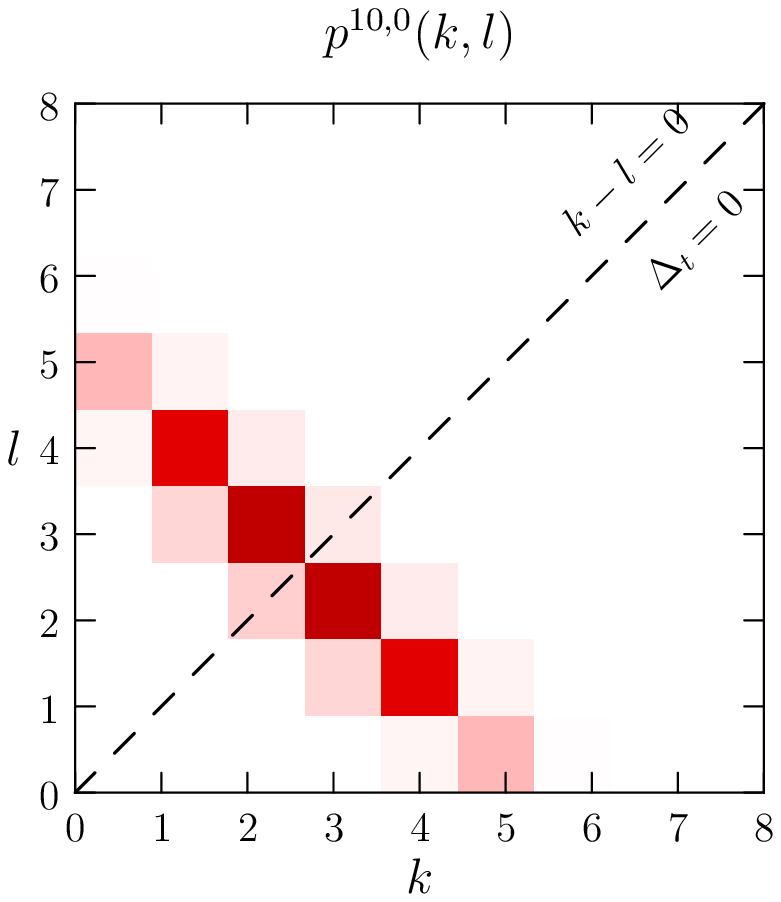}
\\
  \raisebox{4cm}{e)}\includegraphics[width=3.8cm]{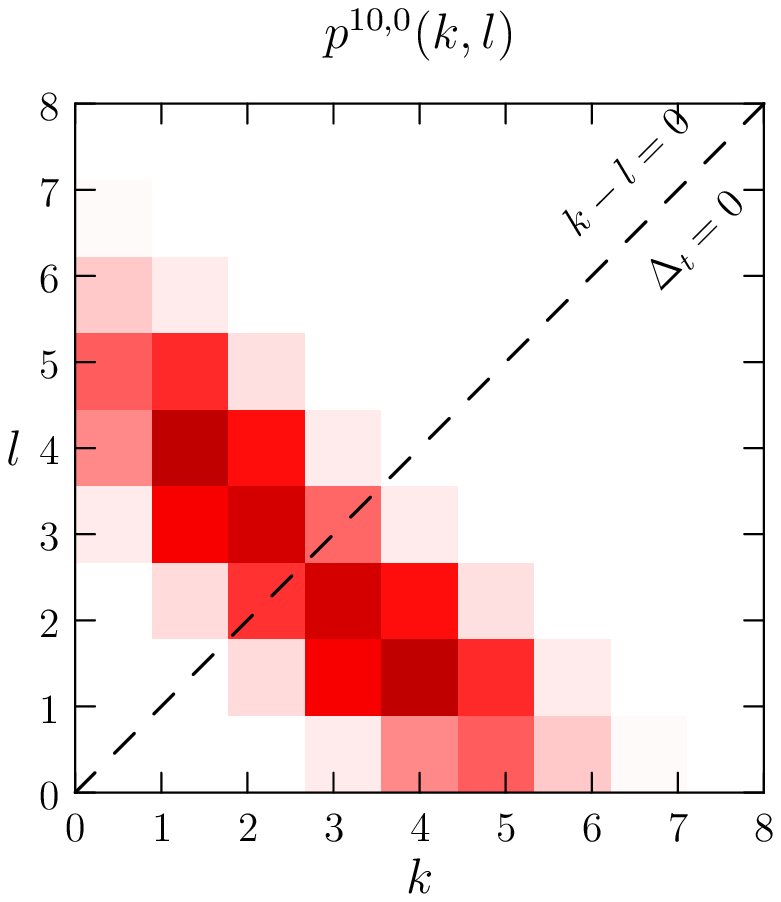}
  \raisebox{4cm}{f)}\includegraphics[width=3.8cm]{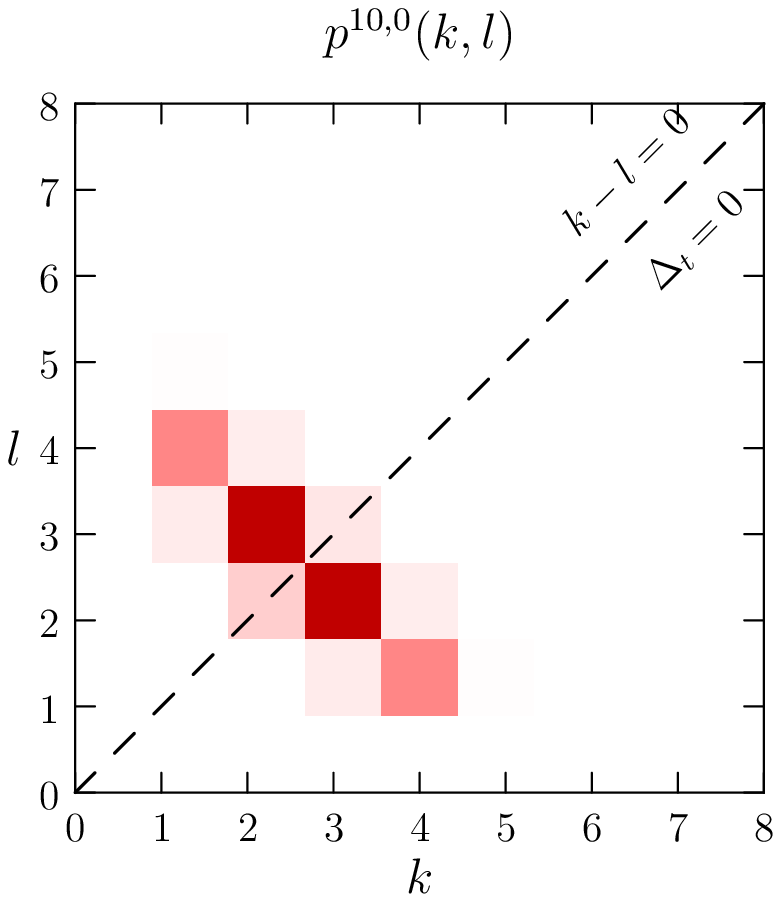}
\end{center}
\caption{The plots of probability distributions $p^{S,\Delta}(\Delta_t)$ numerically computed for the input state (\ref{psi2}) with a total number of photons in range $S_i\in[4,10]$, $r=10\%$, $S=2$, $\Delta=0$ (left column) and $\Delta=2$ (right column). The results were obtained for both perfect (a \& b) and imperfect photodetection, with binomial (c \& d) and Gaussian (e \& f) distribution $d_K(K')$ ($d_L(L')$), representing the detector characteristics. Detailed description is presented in the main text.}
\label{psi2_Si4_10}
\end{figure}

Figs.~\ref{psi2_Si80_120} and \ref{psi2_Si4_10} show similar computations performed for input state (\ref{psi2}) and two ranges of total photon numbers $S_i$: $S_i\in[80,120]$ and $S_i\in[4,10]$, respectively. In the first range we assumed $S=10$ photons registered by the detectors and $\Delta\in\{0,10\}$.  In the second range, total of $S=2$ photons is detected and the difference between the readouts of the detectors is $\Delta\in\{0,2\}$.  Plots of the projection areas a) \& b) represent the case of ideal photon counting detectors, plots c) \& d) -- lossy detectors with efficiency $\eta=80\%$; plots e) \& f) -- imprecise detectors with the Gaussian distribution of the standard deviation $3\sigma=10$ (for $S=10$) and $3\sigma=0.6$ (for $S=2$).

\begin{figure}
\begin{center}
  \raisebox{3cm}{a)}\includegraphics[width=4.5cm]{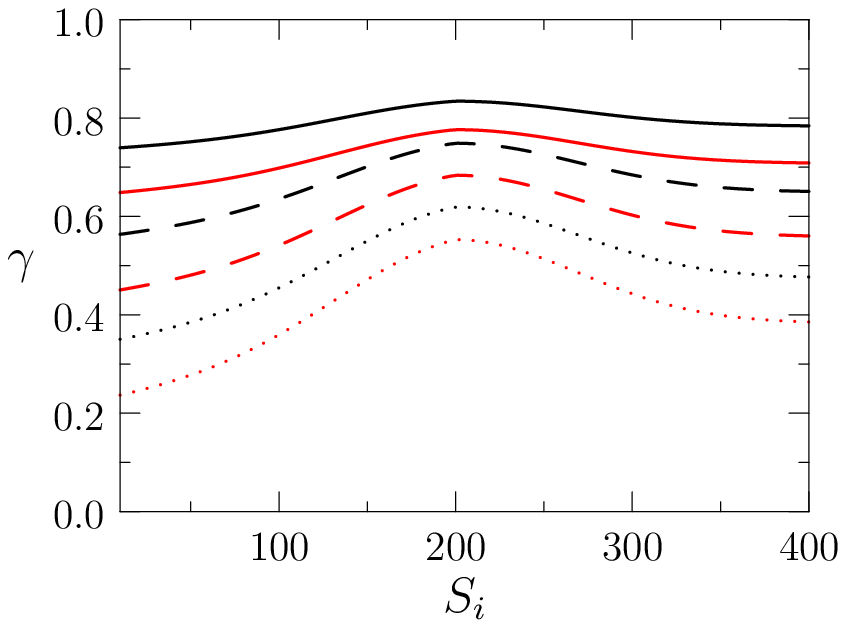}\quad
  \raisebox{3cm}{b)}\includegraphics[width=4.5cm]{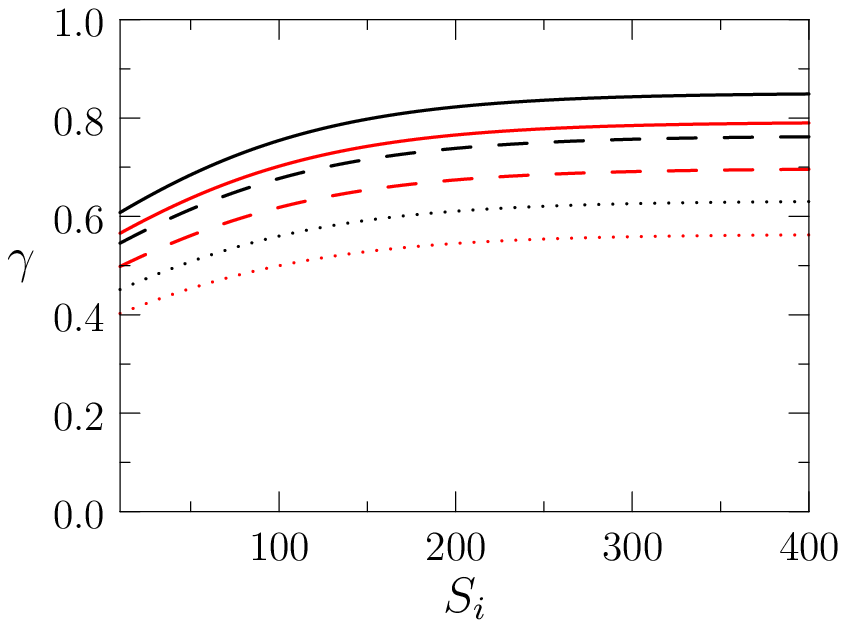}
\end{center}
\caption{The plots of the purity of the state $|\psi_t\rangle$, for the input state (\ref{psi1}) with a total number of photons in range $S_i\in[10,400]$, $r=10\%$, $\Delta=0$, numerically computed for $S=20$ (left figure) and $S=r\cdot S_i$ (right figure). Black curves correspond to the imperfect detection modeled by a binomial distribution with $\eta=5\%, 10\%, 15\%$ (solid, dashed, dotted), red curves -- Gaussian distribution with $3\sigma=5, 10, 20$ (solid, dashed, dotted). Detailed description is presented in the main text.}
\label{fig:purity}
\end{figure}

\begin{figure}
\begin{center}
  \raisebox{3cm}{a)}\includegraphics[width=4.5cm]{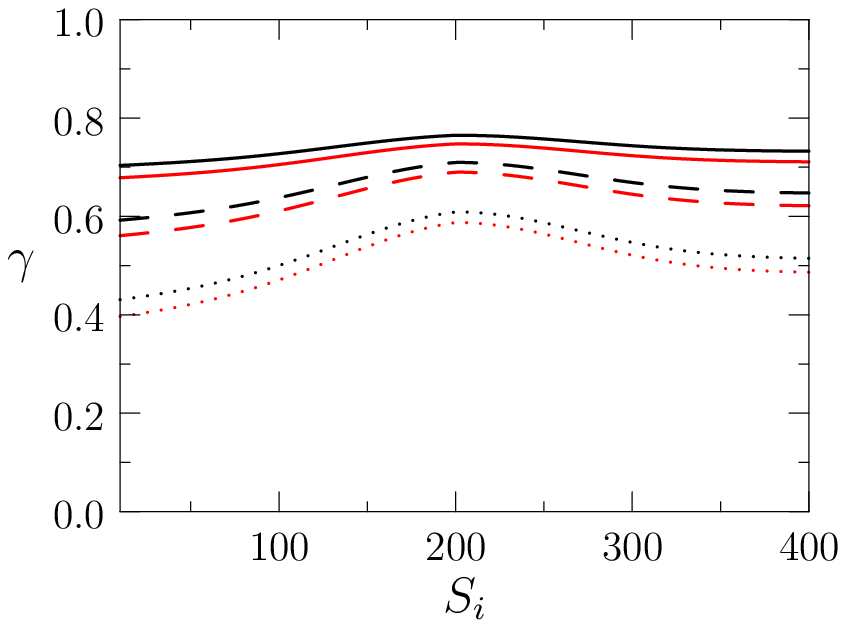}\quad
  \raisebox{3cm}{b)}\includegraphics[width=4.5cm]{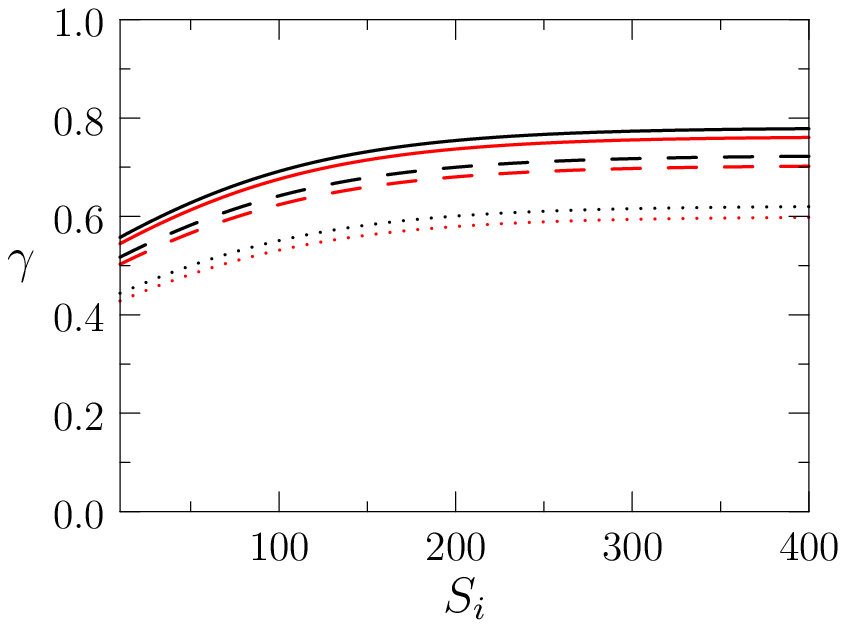}
\end{center}
\caption{The plots of the purity of the state $|\psi_t\rangle$, for the input state (\ref{psi2}) with number of photons in the range $S_{i_1}=0.8\,S_i$, $S_{i_2}=1.2\,S_i$, where $S_i\in[10,400]$, $r=10\%$, $\Delta=0$, numerically computed for $S=20$ (left figure) and $S=r\cdot S_i$ (right figure). Black curves correspond to the imperfect detection modeled by a binomial distribution with $\eta=5\%, 10\%, 15\%$ (solid, dashed, dotted), red curves -- Gaussian distribution with $3\sigma=5, 10, 20$ (solid, dashed, dotted). Detailed description is presented in the main text.}
\label{fig:purity2}
\end{figure}

Finally, Figs.~\ref{fig:purity}-\ref{fig:purity2} depict the purity computed for the states (\ref{psi1})-(\ref{psi2}), respectively, and two models of imperfect detection. The results are presented for two cases, in which imperfect detectors are modeled by binomial (black curves) and Gaussian (red curves) distribution.  Solid black line represents detector efficiency $95\%$, dashed -- $90\%$ and dot-dashed -- $80\%$.  Similarly, red solid line depicts the purity for standard deviation $3\sigma=5$, dashed -- $3\sigma=10$ and dot-dashed -- $3\sigma=20$. For certain values of the parameters (very likely), the purity reaches $80\%$.

The numerical results presented in this Section have shown that the quantum filter executed by the schema in Fig.~\ref{filter}b may be implemented in realistic experimental conditions. It is shown explicitly that the setup preserves its coherent action on an input state even in presence of inefficient photodetection: the photon number distribution of the output state is not distorted and the purity of output state is quite high. Although the results were computed for a specific example of the filtering condition, $\text{C}(S_t,|\Delta_t|)\equiv \{|\Delta_t|\geq \textrm{threshold}\}$, the above conclusions apply to all possible filtering conditions.

\clearpage

\section{Conclusions}

In this paper we have examined experimental scheme which implements a family of quantum non-Gaussian filters. The same setup may apply arbitrary filtering condition $\text{C}(S_t,|\Delta_t|)$ which is set by a relation between the total photon number $S_t$ and the modulus of mode population difference $|\Delta_t|$ in the output state.

Direct applications for some of these filters are already known. It has been shown that filtering according to the condition $\text{C}(S_t,|\Delta_t|)\equiv \{|\Delta_t|\geq \mathrm{threshold}\}$ allows for generation of states useful for quantum optical phase estimation~\cite{Bell2,Demkowicz}. Moreover, this filter helps to increase the distinguishability of macroscopic qubit in analog detection~\cite{MDF}. It also allows for increasing the CHSH-Bell inequality violation by a micro-macro singlet state produced by the phase-covariant quantum cloning~\cite{Bell2}. On the other hand, the condition $\text{C}(S_t,|\Delta_t|)\equiv \{S_t \geq \mathrm{threshold} \}$ allows to increase the generation efficiency of these states~\cite{Vitelli10}. The quantum tasks using the other filtering conditions are not yet known.

All filters work for an arbitrary two-mode input state, pure or mixed, with a small (few photon) or large (mesoscopic) population. We have demonstrated the coherent action of the filter in presence of realistic photodetection involved in the filtering process. The imperfect detection was modeled with a beam splitter of a small reflectivity located in front of a perfect detector and a Weierstrass transform, which implements a Gaussian blur on the unaffected measurement outcome distribution. We have constructed the operators describing the setup of the filter. We have also presented computations for two exemplary initial quantum superpositions, which reveal the structure of the filtered states in the photon number space as well as estimate its purity. 

We believe that the scheme we have discussed will be useful for preparation of the available quantum superpositions for further quantum tasks requiring more complex quantum states than the Gaussian ones.

\section*{Acknowledgments}

This work is supported by the EU 7FP Marie Curie Career Integration Grant No. 322150 ``QCAT'', NCN grant No. 2012/04/M/ST2/00789, FNP Homing Plus project No. HOMING PLUS/2012-5/12, MNiSW co-financed international project No. 2586/7.PR/2012/2 and EU 7FP project BRISQ2 No. 308803.  Computations were carried out at the CI TASK in Gda\'nsk and Cyfronet in Krak\'ow.

\end{document}